\documentclass{article}

\usepackage{graphicx}%
\usepackage{multirow}%
\usepackage{amsmath,amssymb,amsfonts}%
\usepackage{amsthm}%
\usepackage{mathrsfs}%
\usepackage[title]{appendix}%
\usepackage{xcolor}%
\usepackage{textcomp}%
\usepackage{manyfoot}%
\usepackage{booktabs}%
\usepackage{algorithm}%
\usepackage{algorithmicx}%
\usepackage{algpseudocode}%
\usepackage{listings}%
\usepackage{url}
\usepackage{hyperref}

\usepackage[authoryear]{natbib}
\usepackage{multirow} 
\usepackage{makecell} 
\usepackage{rotating} 
\usepackage{subcaption} 
\usepackage{changepage}
\usepackage{xr}
\theoremstyle{theorem}%

\theoremstyle{example}%

\theoremstyle{definition}%

\begin{document}

\title{State Space Modeling of the Ornstein--Uhlenbeck Process with Measurement Error: An Application to Microbiome Data} 

\author{{Shanglun} {Li}, {Toby} {Kenney}, {Hong} {Gu} \\
	Department of Mathematics and Statistics, Dalhousie University.}

\maketitle


\abstract{
Understanding the temporal dynamics of the human microbiome is
critical for deciphering host-microbe interactions and identifying
responses to environmental or physiological disruptions. We develop a
likelihood-based framework to extract  biologically interpretable
inference from microbiome time-series data, based on an Ornstein--Uhlenbeck (OU)
state-space model.
 
We assess temporal stability by using maximum likelihood to estimate
the mean reversion rate of the OU process, and profile likelihood for
confidence intervals. Biologically, higher estimated mean-reversion
rates correspond to stronger regulatory feedback and thus greater
temporal stability, whereas lower rates reflect slower recovery from
perturbations. In the datasets analyzed here, most estimated mean
reversion rates were between 0.12 and 0.36, corresponding to a
half-life in the range 2--6 days for the system to return to the
equilibrium level. These mean-reversion rates are substantially lower
than previous estimates obtained from OU models that did not account
for measurement error.  We also propose an OUSSM to jointly model the
special measurement error covariance structure for log-ratio
transformed data between multiple microbes.

We also propose a statistical testing procedure that compares a
single-parameter model to a split-parameter model across disrupted
time periods.Application to the longitudinal dataset from
\citet{david2014diet} reveals that a considerable fraction of
log-ratio pairs exhibit significant dynamic changes around documented
disruptions, (travel or illness). However, for the majority of pairs
of genera, even though the equilibrium level changed after disruption,
there is not significant evidence of a change to the mean-reversion
rate or diffusion variance.  }

\subsection*{keywords}
Ornstein-Uhlenbeck process; state-space model; profile likelihood confidence interval; microbiome; likelihood ratio test; diagnostic test 



\section{Introduction}
\label{sec1}
Recent advances in high-throughput sequencing have enabled researchers
to investigate the temporal dynamics of the human microbiome with
increasing resolution. Longitudinal microbiome datasets, such as those
from the Human Microbiome Project~\citep{human2012structure} and
subsequent individual-level studies, provide detailed time series of
microbial abundance across various body sites and
individuals. Understanding the temporal behavior of microbial
communities is essential for elucidating ecological stability,
resilience to perturbations, and host-microbiota
interactions~\citep{david2014diet, caporaso2011moving}.

Modeling these microbial time series is challenging due to high
dimensionality, sparsity, compositionality, irregular sampling times,
and the stochastic nature of microbial fluctuations. A common approach
is to focus on log-ratio transformed abundance data, which mitigates
the compositional nature of microbiome profiles. In this context,
stochastic differential equation models, particularly the
Ornstein--Uhlenbeck (OU) process~\citep{uhlenbeck1930theory}, have
gained popularity due to their ability to capture mean-reverting
behavior and equilibrium dynamics. When embedded within a state-space
model (SSM) framework, the OU process allows flexible estimation of
latent microbial trends while accounting for measurement error.

While OU-based SSMs have been widely used for modeling latent
mean-reverting dynamics, most existing applications have focused on
parameter estimation and trajectory reconstruction, with only limited
attention to comprehensive uncertainty quantification. Some studies
report approximate or bootstrap-based confidence intervals for
selected parameters~\citep{dennis2014density, tandeo2011linear,
  michelot2023continuous}, but likelihood-based inference such as
profile-likelihood confidence intervals, formal hypothesis testing,
and systematic diagnostic evaluation appear to be totally absent. In
this work, we develop rigorous likelihood-based uncertainty assessment
and model validation for OU-based latent dynamics.

A particular challenge that arises in the
data from~\citet{david2014diet} is that there were observed external
factors --- illness, travel and dietary changes --- that might have
changed the dynamics of the underlying system, mid-study. Indeed there
was a clear observed difference between the average microbial
abundances before and after these events, violating the stationarity
assumptions of the OU-based SSM (OUSSM).  


While we are not aware of work in the literature that directly tackles
nonstationarity of the OUSSM model, there are several related developments.
%
In movement ecology, \citet{michelot2019state} proposed a
state-switching continuous-time correlated random walk (CTCRW) model
in which the underlying velocity follows an OU process whose
parameters switch according to a continuous-time Markov chain, and
\citet{michelot2023continuous} further extended this idea using
varying-coefficient stochastic differential equations that allow OU
parameters to vary smoothly over time via spline-based covariate
effects. In other fields, \citet{luttinen2014linear} proposed linear
dynamical systems whose transition matrix evolves over time via
weighted combinations of basis matrices. \citet{liu2021gaussian}
developed nonparametric Gaussian-process state-space models with
time-varying inducing points, and \citet{yanchenko2020stanza}
introduced the Stanza model, a nonlinear SSM explicitly designed for
nonstationary latent dynamics. In the discrete-time literature,
\citet{andrews2025inference} studied uniform inference for
autoregressive models with time-varying coefficients, providing
theoretical analogues to inference in piecewise-varying dynamic
systems.

While these studies have addressed smooth or regime-switching forms of
nonstationarity in various contexts, including movement, finance, and
general time-series analysis, they all deal with latent parameter
changes at continuous or unknown times, whereas in our case, there is
a known time at which changes may have occurred. This allows us to
develop more powerful tests using likelihood-based inference.

In summary, this paper develops a comprehensive likelihood-based
framework for modeling longitudinal microbiome dynamics through
Ornstein--Uhlenbeck state-space models (OUSSMs) applied to
abundance-ratio time series. The framework incorporates a structured
measurement-error model for log-ratio transformed microbiome data that
exploits the covariance structure induced by multiplicative sequencing
errors, enables likelihood-based estimation of latent microbial
dynamics, and provides profile-likelihood confidence intervals for
mean-reversion parameters together with formal likelihood-ratio
tests. In particular, we consider two biologically important
hypotheses: (i) whether microbial genera share a common mean-reversion
rate and (ii) whether microbial dynamics remain stable across
disruption-aligned temporal segments. Through simulation studies, we
evaluate the finite-sample performance of the proposed inferential
procedures. We further apply them to the longitudinal microbiome
dataset of \citet{david2014diet}, where documented perturbations such
as travel and illness provide a natural setting for investigating
structural changes in microbial community dynamics.

The remainder of the paper is organized as follows. Section~\ref{OU}
introduces the Ornstein--Uhlenbeck (OU) process and the proposed OU
state-space model (OUSSM), and presents simulation studies
investigating estimation accuracy. Section~\ref{Profile} develops
likelihood-based inference procedures for the OUSSM, including
diagnostic tests, profile-likelihood confidence intervals, and
likelihood-ratio tests. Section~\ref{H_estimate} develops a
measurement-error framework for microbiome log-ratio data and derives
the associated covariance structure induced by multiplicative
sequencing errors. Finally, Section~\ref{data} illustrates the
proposed methods through applications to the longitudinal microbiome
datasets of \citet{david2014diet} and \citet{caporaso2011moving}.

\section{Ornstein-Uhlenbeck Process and OU State Space Model}
\label{OU}
\subsection{OU Process Models}
The OU process satisfies the following linear stochastic differential
equation:
$$dx_t =-\theta(x_t -\mu)dt+\sigma dW_t$$
with the analytical solution for the process given by~\citep{uhlenbeck1930theory}:
$$x_{t_{n+1}} = (1 - e^{-\theta\Delta t_{n+1}})\mu + e^{-\theta\Delta t_{n+1}}x_{t_n}+ \eta_{t_{n+1}}$$
where $\{W_t: t \geq 0\}$ is a Brownian motion, $\Delta t_{n+1}$ is
the time step, $\mu$ represents the mean value towards which the
process tends to revert, $\theta$ represents the rate of mean
reversion, $\sigma^2$ denotes the diffusion variance, $\eta_{t_{n+1}}$
represents the diffusion error with mean 0 and variance
$\sigma^2_{\eta_{t_{n+1}}} = \sigma^2\dfrac{1-e^{-2\theta\Delta
    t_{n+1}}}{2\theta}$.
 
The OU process is widely employed to describe the dynamic behavior of
a continuous-time stochastic process. Initially introduced as a
mathematical framework for understanding Brownian motion and thermal
fluctuations, the OU process has since found extensive applications
across various disciplines, including physics, finance, and
biology. For example, OU processes have been used to model human body
weight~\citep{yatabe2021ornstein}, investment
return~\citep{liang2011optimal}, and the dynamics of infectious
disease spread~\citep{wang2018stochastic}. Recently,
\citet{kenney2020application} have employed an OU process to model
human microbial dynamics and found that the OU process model provides
a significantly better fit to nearly all genera compared to a Brownian
motion model that includes only temporal continuity. Additionally, the
OU model outperforms models that treat time points as independent.

\subsection{OUSSM}
Microbiome time series data are inherently noisy due to both
biological variability and technical artifacts, particularly from 16S
rRNA sequencing. While models such as the OU process are valuable for
capturing features like temporal continuity and mean reversion, they
typically do not account for measurement error --- an omission that
can lead to biased or unstable parameter
estimates. \citet{fox2022development} underscore the substantial
variability in microbial profiles across time and individuals,
pointing to the need for models that can better separate signal from
noise. To address this, we apply the OUSSM, which extends the OU
framework by incorporating a latent layer to explicitly model
measurement error. This allows for more accurate estimation of
microbial dynamics, especially in the presence of technical noise.
The OUSSM is expressed by the following set of equations:
\begin{align*}
y_{t_n} &= \mu + x_{t_n} + \varepsilon \\
x_{t_{n+1}} &= e^{-\theta\Delta t_{n+1}} x_{t_n} + \eta_{t_{n+1}}
\end{align*}
where $x_{t_n}$ is the latent state of the process (for example, the
log-ratio of two microbial abundances) at time $t_n$, and $y_{t_n}$
is the observed value when we attempt to measure $x_{t_n}$. 
%
At the initial time point, $t_0$, the state vector
$x_{t_0}$ follows a normal distribution with mean $a_{t_0}$ and
variance $\sigma^2_{\eta_{t_0}}$. The term $\varepsilon$ represents
the measurement error, assumed to follow a normal distribution with
mean zero and variance $\sigma_{\varepsilon}^2$. $\eta_{t_{n+1}}$
represents the diffusion error with mean 0 and variance
$\sigma^2_{\eta_{t_{n+1}}} = \sigma^2\dfrac{1-e^{-2\theta\Delta
    t_{n+1}}}{2\theta}$. Notably, we have moved the mean value towards
which the process tends to revert ($\mu$) to the observation equation,
simplifying the system. Consequently, the state equation is assumed to
exhibit mean-reverting behavior around zero. The diffusion error
variance at an infinite time point is defined as
$\sigma^2_{\eta_{\infty}} = \dfrac{\sigma^2}{2\theta}$. This
represents the long-term equilibrium value of the variance in the
latent state dynamics.
 
The OUSSM has been successfully applied in various scientific fields,
including the analysis of sea surface temperature
data~\citep{tandeo2011linear} and population
dynamics~\citep{dennis2014density}, particularly in contexts with
irregularly spaced observations and measurement error. However, there
is no published literature on the application of OUSSM to microbiome
data. A unique challenge in this domain is that microbial time series
are typically much shorter than those encountered in climate or
ecological studies. This raises an important question: is the OUSSM
suitable for modeling microbial dynamics when the available time
series is limited in length? In this paper, we aim to investigate this
question and evaluate how reliable inference can be achieved under
such data constraints.

\subsection{OU Process vs. OUSSM}
\label{compare}

In this section, we compare the performance of the OU process and the
OUSSM in estimating the rate of mean reversion, $\theta$, on data
simulated with measurement errors. We simulate time series data under
the OUSSM model. We generate datasets of length $N \in \{100, 300,
500\}$, and true mean reversion rate $\theta\in\{0.1, 0.5, 0.8,
1.0\}$, spanning a range from slow to fast reversion rates. For each
combination of $N$ and $\theta$, we vary the measurement error
variance $\sigma_\varepsilon^2$ from 0 to 5 with an increment of
0.1. Each simulated dataset is generated with a fixed 
stable-state variance $\sigma^2_{\eta_{t_\infty}} = 1$, mean $\mu =
0$, and time step $\Delta t = 1$. For each configuration, 100
replicate data sets are generated and both the OU process and the
OUSSM are fitted.

\begin{figure}[htbp]
\centering
\includegraphics[scale = 0.3]{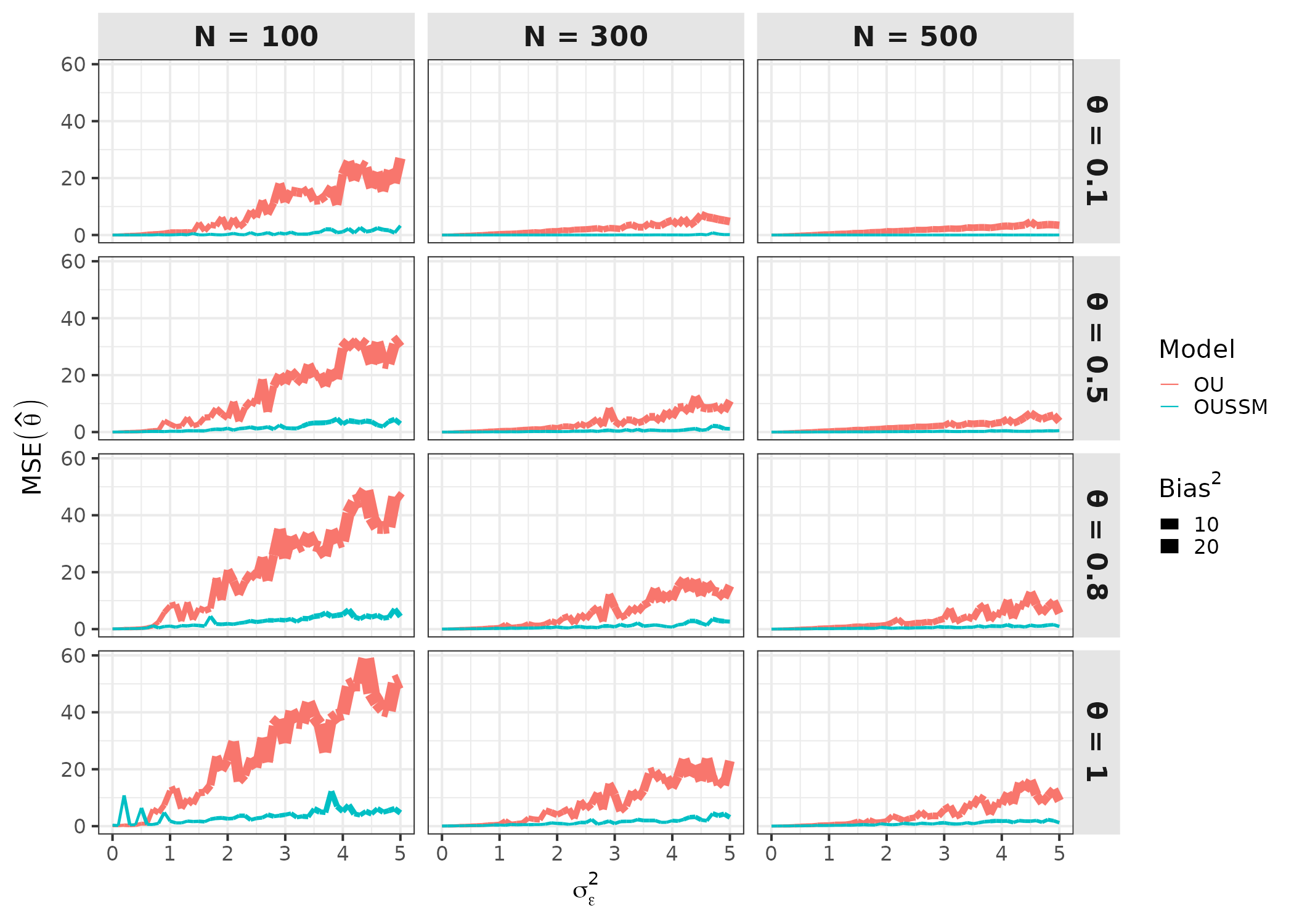}
\caption{ Plots of the mean squared error (MSE) of $\hat{\theta}$
  using the OU model (red line) and the OUSSM (blue line) under
  different scenarios. The scenarios include different ranges of
  measurement error variance, $\sigma^2_{\varepsilon}$, data lengths,
  $N$, and mean reversion rates, $\theta$. The thickness of each line
  represents the squared bias.  }
\label{SSMvsOU}
\end{figure}

Figure~\ref{SSMvsOU} shows the bias and MSE of $\hat{\theta}$ over 100
replicates. As expected, the OU model consistently produces more
biased estimates than the OUSSM. This results in higher MSE, except in
cases when $\sigma_\varepsilon^2$ and $N$ are small, and $\theta$ is
close to 1. In these settings, the bias is smaller, as measurement
error is small, and failing to model measurement error leads to
overestimation of $\theta$, which results in less bias for larger
values of $\theta$. As a result, the increased variability of the
OUSSM causes a larger increase in MSE than the bias from the OU model
in these scenarios.
As the measurement error variance or sample
size increases, however, the advantage of explicitly modeling
measurement error becomes increasingly evident, and the OUSSM
consistently outperforms the OU model.

\subsection{Identifiability and Estimation Accuracy}
\label{Accuracy}

The simulation study in Section~\ref{compare} demonstrates the
advantages of explicitly accounting for measurement error when
estimating the mean-reversion rate. A complementary question is
whether the mean-reversion rate can be estimated accurately across the
full range of scientifically relevant parameter settings. In
particular, the interaction between the mean-reversion rate,
measurement error variance, and observation length may substantially
affect parameter identifiability. To investigate how estimation
accuracy relates to underlying parameter values, we conduct a broader
simulation study.

Because the OUSSM contains multiple parameters, a direct exploration
of the full parameter space would be unfeasible. Fortunately, we can
reduce the necessary exploration by transformation. The full
model contains four parameters,
\[
\Theta=\{\theta,\sigma_\varepsilon^2,\sigma_{\eta_{t_\infty}}^2,\mu\},
\]
However, 
%
defining a rescaled time variable
\(\tilde{t}=\theta t\) sets the mean-reversion rate to \(\theta=1\). A
centered process \(\tilde{x}_t=x_t-\mu\) removes the location
parameter, fixing the long-term mean at \(\mu=0\). Finally, scaling
the state variable by \(\sqrt{2\theta}/\sigma\), such that
\[
\hat{x}_t=\frac{\sqrt{2\theta}}{\sigma}\tilde{x}_t,
\]
normalizes the stationary variance to \(\sigma_{\eta_\infty}^2=1\).

In the discrete-time representation
\[
x_{t_{n+1}} = e^{-\theta \Delta t}x_{t_n}+\eta_{t_{n+1}},
\]
the parameters \(\theta\) and \(\Delta t\) appear only through their
product \(\theta\Delta t\). Consequently, instead of rescaling time so
that $\theta=1$, we may rescale so that \(\Delta t=1\), and interpret
\(\theta\) as the effective mean-reversion rate per observation
interval. Based on these transformations, we can infer the estimation
performance for all parameter values, based on the estimation
performance with $\Delta t=1$, $\mu=0$ and
$\sigma_{\eta_\infty}^2=1$. We therefore fix these values and simulate
data with a range of values for $\theta$ and $\sigma_{\varepsilon}^2$.


We vary \(\log(\theta)\) and \(\log(\sigma_\varepsilon^2)\)
independently from \(-5\) to \(5\) in increments of \(0.1\), resulting
in 101 values for each parameter and 10,201 unique parameter
combinations. For each combination, and for each sequence length
\(N=300\) or \(N=5000\), we simulate 200 datasets and estimate
\(\theta\) using the OUSSM. The mean squared error (MSE) of
\(\log(\hat{\theta})\) is used as the performance measure. This choice
is motivated by the fact that differences on the logarithmic scale
correspond approximately to relative errors on the original scale,
providing a scale-invariant measure of estimation accuracy across a
wide range of values of \(\theta\).

\begin{figure}[htbp]
\centering
\begin{subfigure}{0.495\textwidth}
    \includegraphics[width=\linewidth]{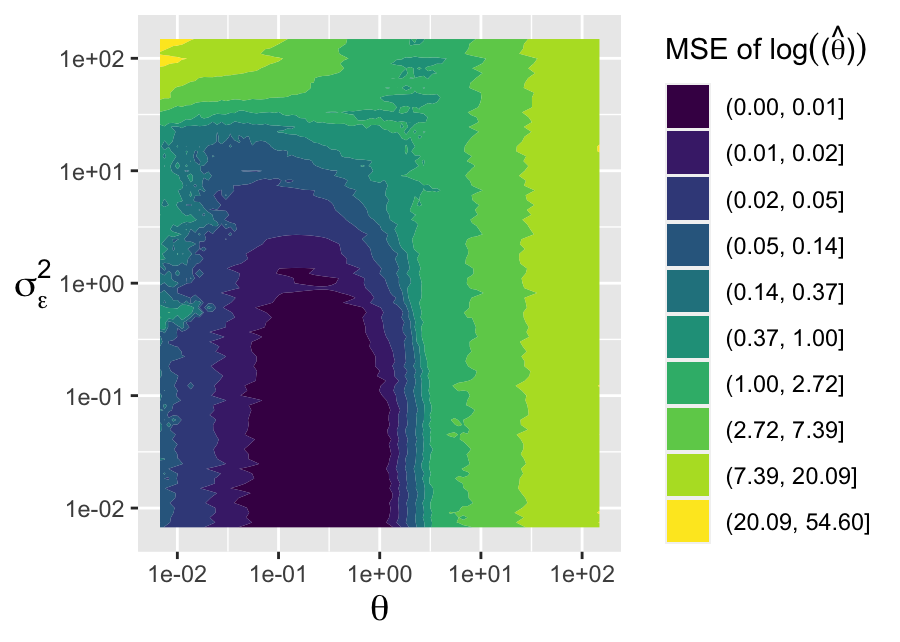}
    \caption{$N=5000$}
    \label{Simulation5000}
\end{subfigure}
\hfill
\begin{subfigure}{0.495\textwidth}
    \includegraphics[width=\linewidth]{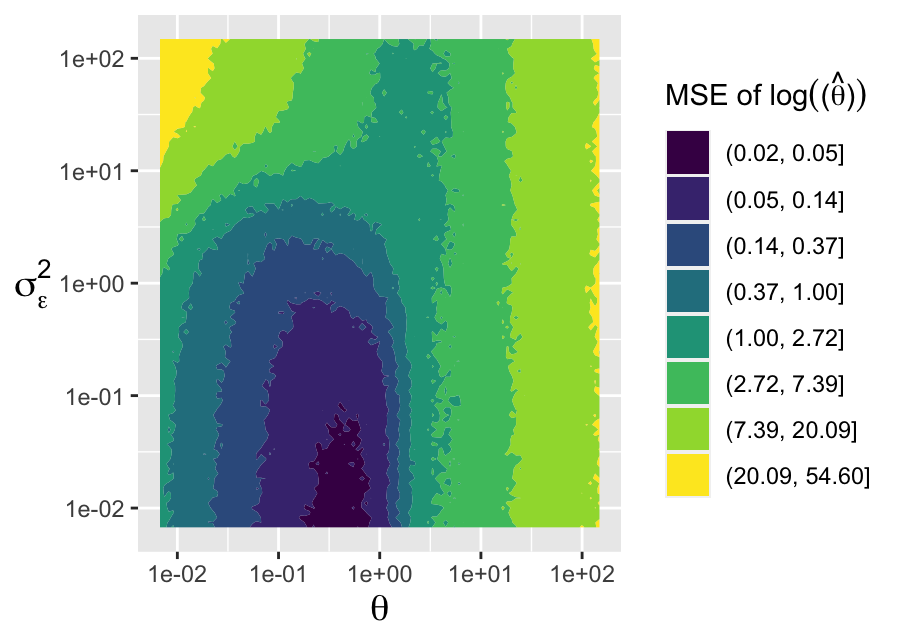}
    \caption{$N=300$}
    \label{Simulation300}
\end{subfigure}
\caption{Contour plots of the MSE of $\log(\hat\theta)$ using the OUSSM. For each of the 101 $\log(\theta)$ and 101 $\log(\sigma_\varepsilon^2)$ values, 200 datasets were simulated and $\hat\theta$ was estimated. (a) $N=5000$, (b) $N=300$.}
\label{SimulationCompare}
\end{figure}

Figure~\ref{SimulationCompare} displays contour plots of the resulting
MSE surface. Darker regions correspond to lower estimation error and
hence greater estimation accuracy. Several features are
evident. First, the surface for \(N=300\) is considerably more
irregular than that for \(N=5000\), reflecting the increased sampling
variability associated with shorter time series. Second, the region of
low estimation error is substantially larger when \(N=5000\),
demonstrating the gains in parameter estimation obtainable from longer
observation sequences.

Nevertheless, both plots exhibit the same overall pattern. Estimation
accuracy is highest when \(\theta\) lies between approximately \(0.1\)
and \(1\) and the measurement-error variance satisfies
\(\sigma_\varepsilon^2 < 1\). When \(\theta\) is very small, the
process exhibits little mean reversion over the observation window,
making it difficult to distinguish \(\theta\) from zero. Conversely,
when \(\theta\) is very large, successive observations become nearly
independent, reducing the information available about the underlying
temporal dynamics. Increasing measurement-error variance further
obscures the latent process and exacerbates these difficulties. As a
result, estimation accuracy deteriorates substantially outside the
region corresponding to moderate mean-reversion rates and relatively
small measurement-error variances.

\section{Inference for the OUSSM Mean Reversion Rate}
\label{Profile}

The simulation studies in Section~\ref{Accuracy} show that estimation
accuracy depends strongly on the sample size and measurement-error
variance. In particular, estimates of the mean-reversion rate
\(\theta\) may exhibit substantial variability when the available time
series is short or the measurement error is large. Consequently, point
estimates alone do not adequately characterize uncertainty.

More generally, despite the attractive theoretical properties of
state-space models, practical estimation can be complicated by model
misspecification, weak identifiability, parameter initialization, and
numerical optimization issues~\citep{auger2016state}. Therefore,
reliable inference requires both accurate parameter estimation and
appropriate uncertainty quantification. In this section, we develop a
likelihood-based inference framework for the OUSSM. We first describe
maximum likelihood estimation and diagnostic procedures for assessing
model adequacy, and then introduce profile likelihood confidence
intervals for the mean-reversion rate \(\theta\).

\subsection{Maximum Likelihood Estimation}
The most commonly employed method for parameter estimation in this
context is the Kalman filter, a recursive algorithm that evaluates the
likelihood through a sequence of one-step-ahead predictions (see, for
example, \citet{durbin2012time}). The log-likelihood function is
$$\log L = -\frac{N + 1}{2}\log 2\pi - \frac{1}{2} \sum_{n=0}^N\left(\log F_{t_n} + \dfrac{v_{t_n}^2}{F_{t_n}} \right)$$
where the terms in the equations are defined as follows:
\begin{align*}
F_{t_n} &= P_{t_n}+\sigma^2_{\varepsilon}\\
v_{t_n}&=y_{t_n}-\mu - a_{t_n}\\
a_{t_{n+1}} &=  e^{-\theta\Delta t_{n+1}} a_{t_n} + e^{-\theta\Delta t_{n+1}}K_{t_n}v_{t_n}\\
K_{t_n} &= \dfrac{P_{t_n}}{P_{t_n}+\sigma^2_{\varepsilon}}\\
P_{t_{n+1}} &=  e^{-2\theta\Delta t_{n+1}}  P_{t_n}(1 - K_{t_n}) + \sigma^2\frac{1- e^{-2\theta\Delta t_{n+1}}}{2\theta}
\end{align*}

The parameters of interest are
$\Theta=\{\theta,\sigma_\varepsilon^2,\sigma^2,\mu\}$.
Since $\theta$, $\sigma_\varepsilon^2$, and $\sigma^2$ are strictly positive,
maximum likelihood estimation was performed on the transformed parameter scale
$(\log\theta,\log\sigma_\varepsilon^2,\log\sigma^2,\mu)$,
yielding an unconstrained optimization problem.

For each dataset, the exact Gaussian likelihood was evaluated using the Kalman filter, with the latent state initialized from its stationary distribution, whose variance is $\sigma^2/(2\theta)$. The negative log-likelihood was minimized using the Nelder--Mead algorithm implemented in R's \texttt{optim} function. Starting values were chosen using simple data-driven rules: $\theta^{(0)}=0.1$, $\mu^{(0)}$ equal to the sample mean, and
\[
\sigma^{2(0)}
=
\sigma_\varepsilon^{2(0)}
=
\widehat{\mathrm{Var}}(y_{t_n}-y_{t_{n-1}})/3.
\]
The variance initialization is motivated by the approximation
\[
\mathrm{Var}(\Delta y)
=
2\sigma_\varepsilon^2
+
\sigma^2
\frac{1-e^{-\theta}}{\theta}
\approx
2\sigma_\varepsilon^2+\sigma^2,
\]
which holds for unit sampling intervals and moderate values of $\theta$.

Because state-space models can exhibit weak identifiability and
numerical estimation difficulties~\citep{auger2016state}, we assessed
convergence by screening for degenerate solutions after
optimization. Fits were flagged when the estimated diffusion variance
collapsed to a near-boundary value ($\hat{\sigma}^2 < e^{-10}$),
rendering the OU component effectively absent, or when the estimated
mean-reversion rate was implausibly large or small
($|\log\hat{\theta}| \ge 7$). Optimization was initialized using a
single data-driven starting value for each fit, without random
restarts. Across all simulation settings, approximately 1.7\% of fits
were flagged, indicating that convergence or identifiability problems
were relatively uncommon. Flagged fits were excluded when computing
the MSE of $\hat{\theta}$, but retained for analyses based on
$\log(\hat{\theta})$, which are less sensitive to extreme estimates.

The implementation is available as an R package at
\url{https://github.com/ShanglunLi/OUSSM1d}.

\subsection{Testing Model Assumptions and Mean Reversion under OUSSM}
\label{diag_test}

The Ornstein--Uhlenbeck State-Space Model (OUSSM) relies on several
key assumptions: (i) the process and measurement errors are
independent, Gaussian, and serially uncorrelated; (ii) the latent
process is stationary, fluctuating around a long-term mean; and (iii)
the state and observation equations are linear with constant
parameters over time. To assess whether these assumptions hold and to
verify that the microbial time series exhibit mean-reverting dynamics,
we conduct two diagnostic analyses: a residual-based test for
distributional adequacy, and a likelihood ratio test for mean
reversion.

\subsubsection{Normality of residuals}
To evaluate whether the OUSSM adequately captures the distributional
structure of the data, we examine the one-step-ahead standardized
prediction residuals from the Kalman filter:
\[
e_{t_n} = \frac{v_{t_n}}{\sqrt{F_{t_n}}},
\]
where $v_{t_n} = y_{t_n} - \mu - a_{t_n}$ is the one-step-ahead
prediction error and $F_{t_n} = P_{t_n}+\sigma^2_{\varepsilon}$ is the
corresponding prediction variance. Under correct model specification,
the residuals $\{e_{t_n}\}$ should follow a standard normal
distribution with mean 0 and variance 1. We assess this assumption
formally via the Shapiro--Wilk test for normality. Systematic
deviations from normality may indicate model misspecification, such as
nonlinear dynamics, non-Gaussian process noise, or heteroskedastic
measurement error.

\subsubsection{Independence of residuals}  
In addition to normality, the Kalman filter assumes that the
standardized one-step-ahead prediction errors $\{e_{t_n}\}$ are
serially uncorrelated. We assessed this assumption using the
Ljung--Box test for residual autocorrelation. Because the sampling
intervals in microbiome time series are irregular, we adapted the test
by mapping the observation times onto a regularized grid and treating
unobserved time points as missing samples. The test was then applied
to the one-step-ahead prediction errors as a heuristic diagnostic for
independence. Models with non-significant test results ($p > 0.05$)
were considered to satisfy the white-noise assumption required by the
Kalman filter.

In the context of microbiome data, the independence between process
and measurement errors is also biologically well justified. The
process noise $\eta_{t_n}$ represents intrinsic stochastic
fluctuations in microbial abundances, whereas the measurement noise
$\varepsilon_{t_n}$ arises from independent technical factors such as
sampling variability, DNA extraction, and sequencing depth. Because
these two sources of randomness occur at distinct stages of data
generation and have no causal interaction, their independence is a
reasonable assumption. Any remaining serial dependence in the
one-step-ahead prediction errors would instead indicate model
misspecification, such as unmodeled dynamic components or correlated
observation noise.

\subsubsection{Stationarity and mean reversion}
A core assumption of the OUSSM is that the latent process is
stationary, meaning it reverts to a stable long-term mean $\mu$ with
mean-reversion rate $\theta > 0$. Formally, this is equivalent to
requiring $|\phi| = e^{-\theta \Delta t} < 1$ in the discrete-time
representation. Classical unit-root tests, such as the Dickey-Fuller
test, are commonly used to test this stationarity condition in
autoregressive models. However, in our framework, this property is
embedded within the OUSSM itself: $\theta$ is explicitly estimated via
maximum likelihood, and testing whether $\theta = 0$ versus $\theta >
0$ directly assesses mean reversion.

Accordingly, we perform a log-likelihood ratio test comparing the full
OUSSM to a nested baseline model --- the local level model --- which
corresponds to the special case $\theta = 0$:
\begin{align*}
y_{t_n} &= \mu + x_{t_n} + \varepsilon_{t_n}, \\
x_{t_{n+1}} &= x_{t_n} + \eta_{t_{n+1}},
\end{align*}
where $\varepsilon_{t_n} \sim \mathcal{N}(0, \sigma^2_{\varepsilon})$ and $\eta_{t_{n+1}} \sim \mathcal{N}(0, \sigma^2 \Delta t_{n+1})$. The hypotheses are:
\begin{align*}
H_0 &: \theta = 0 \quad \text{(no mean reversion / non-stationary dynamics)}, \\
H_A &: \theta > 0 \quad \text{(mean-reverting / stationary dynamics)}.
\end{align*}
Because the local level model is not a regular special case of the OUSSM, the null distribution of the log-likelihood ratio statistic is non-standard. We therefore approximate it using a parametric bootstrap under $H_0$.

\subsubsection{Linearity and parameter constancy}
The OUSSM further assumes that both the state and observation
equations are linear with constant parameters over time. While no
separate formal test is conducted for this assumption, violations
typically manifest as systematic patterns or changing variance in the
standardized residuals. For instance, nonlinear latent dynamics or
time-varying parameters can produce departures from normality or
residual autocorrelation, which are captured by the diagnostics
above. Thus, the residual analyses indirectly assess the adequacy of
the linear, time-homogeneous structure assumed by the OUSSM.

Together, these diagnostics address the core OUSSM assumptions. The
residual-based tests validate the Gaussian distribution and temporal
independence of the innovation sequence, while the likelihood ratio
test formally evaluates the presence of mean reversion and hence the
stationarity assumption. Since the test of $\theta = 0$ versus $\theta
> 0$ directly evaluates the same condition that underlies classical
unit-root tests, no additional formal stationarity test (e.g.,
Dickey--Fuller) is required. These combined diagnostics confirm both
the adequacy of the model specification and the mean-reverting nature
of microbial dynamics when supported by the data.

\subsection{Profile Likelihood Confidence Intervals}
In this section, we explore two methods for the inference of the MLE,
$\hat\theta$. Two types of confidence intervals are compared: the
confidence interval based on asymptotic likelihood theory and the
confidence interval based on profile likelihood.
 
From standard likelihood theory, MLEs asymptotically follow a normal
distribution with mean equal to the true parameter value and
covariance given by the inverse of the Fisher information
matrix. Thus, the marginal distribution of $\log\hat{\theta}$ is
approximately normal, and the $(1 - \alpha) \times 100\%$ confidence
interval for $\log\theta$ is estimated as
\[
\log(\hat{\theta}) \pm z_{\alpha/2} \sqrt{(H(\hat{\Theta})^{-1})_{1,1}}
\]
where $H(\hat{\Theta})$ is the observed negative Hessian matrix
evaluated at the MLE $\hat{\Theta}$, $(H(\hat{\Theta})^{-1})_{1,1}$ is
the diagonal entry of the matrix $H(\hat{\Theta})^{-1}$ corresponding
to $\log(\hat\theta)$, and $z_{\alpha/2}$ is the standard normal
critical value.
 
However, it is important to note that the negative Hessian matrix can
occasionally become nearly singular, which may result in instability
when calculating the inverse matrix. This instability can lead to
potential issues with the inference.
 
The profile likelihood for the mean reversion parameter $\theta$ is
obtained by maximizing the log-likelihood function over all nuisance
parameters $(\sigma^2_\eta, \sigma^2_\varepsilon, \mu)$ for each fixed
value of $\theta$. In practice, we evaluate this function over a grid
of $\theta$ values centered around the MLE, $\hat{\theta}$, on a
logarithmic scale. At each grid point, $\theta$ is held fixed while
the remaining parameters are re-estimated via numerical optimization
of the Kalman filter log-likelihood. This yields a smooth profile
likelihood curve $\ell_p(\theta) = \max_{\sigma^2_\eta,
  \sigma^2_\varepsilon, \mu} \ell(\theta, \sigma^2_\eta,
\sigma^2_\varepsilon, \mu)$.

The $100(1-\alpha)\%$ confidence interval is then defined as the set
of $\theta$ values satisfying
\[
2\left[\ell_p(\hat{\theta}) - \ell_p(\theta)\right] \le \chi^2_{1,\,1-\alpha},
\]
that is, all values of $\theta$ whose profile log-likelihood lies
within $\tfrac{1}{2}\chi^2_{1,\,1-\alpha}$ of the maximum. In
implementation, we compute this numerically by locating the grid
points where $\ell_p(\theta)$ first falls below the cutoff
$\ell_p(\hat{\theta}) - \tfrac{1}{2}\chi^2_{1,\,1-\alpha}$. This
procedure yields finite-sample confidence intervals with correct
coverage.
 
\subsection{Coverage Simulations}

\begin{center}
\begin{table}[h]
\centering
\begin{tabular}{l@{\hspace{1cm}}llll}
\hline\hline
$N$ & \ 100 & \ 500 & 1000 & 5000  \\\hline\hline
&\multicolumn{4}{c} {Coverage Rate} \\\hline
Asymptotic Likelihood  & 97\% & 82\% & 86\% & 95\% \\
Profile Likelihood & 93\% & 96.5\% & 95\% & 94\% \\\hline
&\multicolumn{4}{c} {Median of Interval Width} \\\hline
Asymptotic Likelihood  & 3.40 & 1.97 & 1.47 & 0.72 \\
Profile Likelihood & 4.61 & 1.74 & 1.34 & 0.67 \\\hline
&\multicolumn{4}{c} {IQR of Interval Width} \\\hline
Asymptotic Likelihood  & 3.05 & 1.05 & 0.47 & 0.11 \\
Profile Likelihood & 1.69 & 0.69 & 0.47 &  0.13 \\\hline
\end{tabular}
\caption{Three metrics of the confidence interval: confidence interval coverage rate of true $\log(\theta)$, median confidence interval width, and IQR of confidence interval width. The parameters of the model are fixed at $\theta = 1$, $\sigma_\varepsilon^2 = 1$, $\sigma^2_{\eta_{\infty}} = 1$, and $\mu = 0$. The number of replicate data sets is 200.}
\label{table:1}
\end{table}
\end{center}

Table~\ref{table:1} presents a comparison of confidence interval
properties for $\log(\theta)$ obtained from the asymptotic likelihood
and profile likelihood methods. The parameters of the model are fixed
at $\theta = 1$, $\sigma_\varepsilon^2 = 1$, $\sigma^2_{\eta_{\infty}}
= 1$, and $\mu = 0$, with 200 simulated data sets. The results show
that the profile likelihood method generally attains coverage close to
the nominal level across all sample sizes, whereas the coverage rate
for the asymptotic likelihood method is too small for moderate $N$. In
terms of interval width, the differences are small. Profile likelihood
intervals are somewhat wider when $N=100$, but become narrower and
more consistent as $N$ increases. Overall, the profile likelihood
provides a more reliable inference procedure for $\theta$, especially
when the sample size is limited or moderate.

\section{Measurement Error Structure for Microbiome Log-Ratio Data}
\label{H_estimate}
\subsection{Multiplicative Error Framework}

In microbiome sequencing data, measurement error arises from several
sources, including sampling variation, sequencing depth differences,
and amplification bias. These factors primarily act as scaling
effects, making a multiplicative formulation more appropriate than an
additive one. Formally, we assume that the observed OTU count
$c_{t_n}^{(g)}$ for genus $g$ at time $t_n$ arises from the underlying
true abundance $\tilde{c}_{t_n}^{(g)}$ through a multiplicative
measurement error:
\[
c_{t_n}^{(g)} = \tilde{c}_{t_n}^{(g)}S_{t_n}a^{(g)} \varepsilon_{t_n}^{(g)}
\]
where $S_{t_n}$ is the sequencing depth at time point $t_n$, $a^{(g)}$
is the average amplification bias for genus $g$ and
$\varepsilon_{t_n}^{(g)}$ is a multiplicative error term that
incorporates random factors, and can be assumed independent for
different genera and/or time points. This multiplicative formulation
respects the strictly positive nature of count data and reflects the
proportional scaling inherent to sequencing processes, in which deeper
sequencing or amplification efficiency inflates all counts by a
roughly constant factor. In contrast, an additive-error model could
yield negative values and would not appropriately represent the
scaling behavior observed in sequencing experiments. Similar
multiplicative error formulations are implicit in standard
compositional data frameworks~\citep{aitchison1982statistical},
sequencing bias models~\citep{mclaren2019consistent}, and
measurement-error studies in microbiome
research~\citep{llorens2021benchmarking, brill2022testing}.

\subsection{Error Covariance Structure}

We employ a log-ratio transformation approach to handle the OTU counts
data. Specifically, for each pair of abundant genera $i$ and $j$,
\[
y_{t_n}^{(i,j)} = 
\log\!\left(
\frac{c_{t_n}^{(i)}}{
c_{t_n}^{(j)} }
\right),
\]
where $c_{t_n}^{(g)}$ is the OTU count of genus $g$. Hence, for each
body site of each subject, we obtain an $n_a \times n_a$ matrix of
time series excluding the diagonal elements, where $n_a$ denotes the
number of abundant genera retained for analysis. Each matrix entry
$(i, j)$ corresponds to a sequence of log-ratios $\{y_{t_n}^{(i,j)}\}$
across time points. Each log-ratio time series is analyzed
independently using the OUSSM framework to obtain parameter estimates
$\hat{\Phi} = \left\{\hat{\theta}, \hat{\mu},
\hat{\sigma}^2_\varepsilon, \hat{\sigma}^2\right\}$. Therefore, for
each body site of each subject, these estimates are organized into
four $n_a \times n_a$ matrices --- one for each parameter --- corresponding
to the set of abundant genera considered. Specifically, let
$\hat{{\Theta}} = [\hat{\theta}_{ij}], \hat{{M}} = [\hat{\mu}_{ij}],
\hat{{H}} = [\hat{\sigma}^2_{\varepsilon,ij}], \hat{{\Sigma}} =
    [\hat{\sigma}^2_{ij}]$, where each entry $(i,j)$ contains the
    estimate derived from the log-ratio time series between genera $i$
    and $j$ with diagonal entries NA.

After applying the log-ratio transformation between two genera~$i$
and~$j$, the originally multiplicative errors become additive on the
log scale:
\[
y_{t_n}^{(i,j)} = 
\log\!\left(\frac{c_{t_n}^{(i)}}{c_{t_n}^{(j)}}\right)
= 
\log\!\left(\frac{\tilde{c}_{t_n}^{(i)}}{\tilde{c}_{t_n}^{(j)}}\right)+ \log\!\left(\frac{a^{(i)}}{a^{(j)}}\right)
+ \log\!\left(\frac{\varepsilon_{t_n}^{(i)}}{\varepsilon_{t_n}^{(j)}}\right) = \mu^{(i,j)} +x^{(i,j)}_{t_n} + a^{(i,j)} + \varepsilon^{(i,j)}_{t_n} 
\]
where $\mu^{(i,j)}$ denotes the equilibrium mean of the mean reverting
process for the log-ratio time series of the genus pair $(i,j)$,
$x^{(i,j)}_{t_n}$ represents the latent OU process for the pair,
$a^{(i,j)}$ represents the constant bias term arising from systematic
differences in amplification efficiency between genera $i$ and $j$,
and $\varepsilon^{(i,j)}_{t_n}$ is the associated measurement error
with
$\log\!\left(\dfrac{\varepsilon_{t_n}^{(i)}}{\varepsilon_{t_n}^{(j)}}\right)
= \varepsilon^{(i,j)}_{t_n} \sim N(0, H_{ij} = \sigma^2_{\varepsilon,
  ij})$.

Each measurement error log-ratio can be expressed as the difference
between genus-total log errors,
\[
\log\!\left(\frac{\varepsilon_{t_n}^{(i)}}{\varepsilon_{t_n}^{(j)}}\right) = \log\!\left(\varepsilon_{t_n}^{(i)}\right) - \log\!\left(\varepsilon_{t_n}^{(j)}\right) \sim N(0, \sigma^2_{\varepsilon, i} + \sigma^2_{\varepsilon, j})
\]
where $\sigma_{\varepsilon,g}^2$ is the variance of
$\log\big(\varepsilon_{t_n}^{(g)}\big)$. Hence, the measurement-error
variance for any pairwise log-ratio satisfies
\[
H_{ij} = H_{i} + H_{j}, \quad \text{for all } i \neq j.
\]
where $H_{g}$ represents the variance of
$\log\big(\varepsilon_{t_n}^{(g)}\big)$. This implies that any
off-diagonal element of $H$ can be expressed as the sum of two values
corresponding to the row and column respectively. This assumption
allows us to reparametrize the matrix $H$, significantly reducing the
number of parameters. The amplification bias terms $a^{(i,j)}$ are the
same for all time points, so affect the estimated mean of the OU
process, but not the other parameters.

\subsubsection{Estimation Procedure}
To exploit this structure, we proceed in two stages. First, we obtain
initial estimates of the measurement-error variances $\hat H = [\hat
  H_{ij}]$ from the OUSSM fits to the pairwise log-ratio time series
between genera $i$ and $j$. At this stage, the goal is only to obtain
unconstrained estimates of $\hat H$ from the individual OUSSM fits;
the other model parameters, $\{\hat\Theta,\hat M,\hat\Sigma\}$, are
temporarily disregarded. These unconstrained estimates are then used
to infer the values $\hat H_{i}$ by solving a linear regression
problem derived from the additive relationship $H_{ij} = H_{i} +
H_{j}$.
 
Formally, the constraints can be written in vectorized form as
\[
{h} = A {h}_{\text{genus}} + {\varepsilon}, \quad {\varepsilon} \sim N({0}, \sigma^2 I),
\]
where ${h}= \left(\hat H_{12}, \hat H_{13}, \hdots, \hat
H_{(n_a-1)n_a} \right)^T$ is the vector collecting all estimated upper
triangular off-diagonal entries, ${h}_{\text{genus}} = (H_{1}, H_{2},
\dots, H_{n_a})^T$ is the vector of variances for each genus, and $A
\in \mathbb{R}^{m \times n_a}$, where $m = \binom{n_a}{2}$ and each
row has ones in columns $i$ and $j$ (for the row corresponding to
$H_{ij}$), and zeros elsewhere:
\[
A_{(ij),k} = \begin{cases}
1, & k = i \text{ or } k = j, \\
0, & \text{otherwise}.
\end{cases}
\]
 
Solving the above linear regression problem using ordinary least
squares gives:
\[
\hat{{h}}_{\text{genus}} = (A^T A)^{-1} A^T {h}.
\]
 
Note that
$$(A^TA)_{kl}=\sum_{i<j}A_{(ij),k}A_{(ij),l}=\left\{
\begin{array}{ll}
\sum_{i<j}1_{i=k,j=l}+1_{i=l,j=k}&\textrm{if }k\ne l\\
\sum_{i<j}1_{i=k}+1_{j=k}&\textrm{if }k=l\end{array}\right.
=\left\{
\begin{array}{ll}
1&\textrm{if }k\ne l\\
n_a-1&\textrm{if }k=l\end{array}\right.
$$
so $A^TA=(n_a-2)I+11^T$, giving
$(A^TA)^{-1}=\frac{1}{(n_a-1)(n_a-2)}\left((n_a-1)I-\frac{11^T}{2}\right)$ and
thus $h_{\textrm{genus},i}=\frac{\sum_{j\ne i}H_{ij}}{(n_a-2)}-\frac{\sum_{j\ne k}H_{jk}}{(n_a-1)(n_a-2)}$.

Using $\hat{{h}}_{\text{genus}}$, we reconstruct the full symmetric
measurement error variance matrix:
\[
\hat{H}_{ij} = \begin{cases}
\text{NA}, & \text{if } i = j, \\
\hat{H}_{i} + \hat{H}_{j}, & \text{if } i \neq j.
\end{cases}
\]

The off-diagonal elements $\hat{H}_{ij}$ are subsequently treated as
fixed measurement-error variances for the corresponding genus-pair
analyses.

The two-stage approach described above enforces internal consistency
among genus pairs and substantially reduces the effective number of
free parameters, leading to improved numerical stability and lower
estimation variability for the remaining model
parameters. Conceptually, one could instead formulate a single joint
model that simultaneously fits all log-ratio series and estimates a
fully constrained $H$, but such an approach would require
high-dimensional state-space optimization, which is computationally
demanding for realistic numbers of genera. The proposed two-stage
procedure therefore provides a practical compromise that retains the
essential covariance structure implied by the multiplicative
measurement error formulation while remaining computationally
efficient.

The structured measurement-error model developed above provides
estimates of the log-ratio measurement-error variances that are
subsequently treated as known in the microbiome analyses that
follow. We now apply the proposed methodology to investigate temporal
stability and disruption-induced changes in longitudinal gut
microbiome data.

\subsection{Single-Genus Log-Ratio Representation}
\label{single}

We also perform single genus analyses, where we analyze the log-ratio
of one genus abundance over the total abundance of all observed
microbes.
\[
y_{t_n}^{(i,i)} = \log\!\left(\frac{c_{t_n}^{(i)}}{\sum_{g} c_{t_n}^{(g)}}\right).
\]
This quantity reflects how the abundance of genus~$i$ changes relative
to the overall microbial community. Because the denominator aggregates
measurement noise from all taxa, both terms are influenced by shared
fluctuations in sequencing depth and amplification bias. As a result,
random errors accumulate and propagate through the ratio, increasing
its variability and amplifying the effective impact of measurement
error on the estimated dynamics. Despite this increased sensitivity to
noise, the single-genus log-ratio remains biologically intuitive, as
it directly captures changes in the relative dominance of a single
taxon within the community over time.  Unlike pairwise log-ratios,
this representation yields one time series per genus and therefore
provides a natural basis for joint modelling of all genera
simultaneously.

\section{Microbiome Applications}
\label{data}

In this section, we examine the dataset of \citet{david2014diet} and
the moving picture dataset~\citep{caporaso2011moving} to investigate
the temporal dynamics of the human gut microbiome.

\subsection{Data Description and Preprocessing}\label{preprocessing}

The moving picture dataset~\citep{caporaso2011moving} comprises
longitudinal microbiome samples collected from two healthy individuals
across multiple body sites, including the gut, tongue, left hand and
right hand. The sampling was conducted over a period of 6 to 15
months, with high temporal resolution samples collected almost daily,
though some gaps of several days occurred. This dataset captures
natural microbiome fluctuations in a stable environment. Sequencing
targeted the V4 region of the 16S rRNA gene using Illumina HiSeq
technology. For the Moving Picture dataset, following
\citet{kenney2020application}, we restricted our analysis to abundant
genera with total Operational Taxonomic Unit (OTU) counts greater than
20,000 across all samples within each body site for a given
individual.
 
The dataset of \citet{david2014diet} also samples longitudinal gut
microbiome data from two individuals over a period of approximately
one year. This dataset includes a dietary intervention and samples
were collected less frequently than in the Moving Picture dataset,
typically every 1--2 days. Although both gut and tongue samples were
collected, our analysis is restricted to gut samples to ensure
comparability across subjects. Sequencing in this dataset targeted the
V2 region of the 16S rRNA gene using the Roche 454 platform. For the
dataset of \citet{david2014diet}, following
\citet{kenney2020application}, we restricted our analysis to abundant
genera with total counts across all samples for one subject exceeding
10,000 and with an average proportion greater than 0.005 for that
subject. Here, the proportion refers to the relative abundance of each
genus, calculated as the number of reads assigned to that genus
divided by the total number of reads in the
sample. Table~\ref{Data_des} summarizes the number of observations and
the number of abundant genera for each individual and body site across
the Moving Picture and David {\it et
  al}\ datasets~\citep{david2014diet,caporaso2011moving}.

\begin{table}[ht]
\centering
\begin{tabular}{cl|cccc|c}
\hline
 & & \multicolumn{4}{c|}{ Moving Picture} &  David {\it et al.}\ \\
 & & Gut & Tongue & R palm & L palm & Gut \\
\hline
\multirow{2}{*}{\makecell[c]{No. of\\Observations (N)}} 
  & Person 1 & 131 & 135 & 134 & 134 & 342 \\
  & Person 2 & 336 & 373 & 359 & 365 & 192 \\
\hline
\multirow{2}{*}{\makecell[c]{No. of Abundant\\  Genera ($n_a$)}} 
  & Person 1 & 12 & 11 & 8 & 31 & 16 \\
  & Person 2 & 32 & 19 & 64 & 30 & 12 \\
\hline
\end{tabular}
\caption{The number of observations and abundant genera for each individual and body site.}
\label{Data_des}
\end{table}
 
An issue with log-transformed analysis of microbiome data is that
microbiome data often includes zero counts, for which the logarithm is
undefined ($-\infty$). Because we restrict our focus to the most
abundant genera, the incidence of zeros is low (see Table
\ref{Data_zero}). Following \citet{kenney2020application}, zero counts
were replaced with a pseudocount of 0.3 to enable log-scale
transformation. Although \citet{kenney2020application} did not provide
a formal justification for the value 0.3, the rarity of zeros among
the abundant genera means that adding a small constant provides
numerical stability with minimal bias. Similar ad-hoc pseudocount
replacements are widely used in microbiome log-ratio analyses to avoid
discarding samples and to preserve comparability across
taxa~\citep{kaul2017analysis, lin2020analysis}.

\begin{table}[ht]
\centering
\begin{tabular}{clcccc|c}
\hline
 & & \multicolumn{4}{c|}{ Moving picture} &  David {\it et al.} \\
 & & Gut & Tongue & R palm & L palm & Gut \\
\hline
\multirow{2}{*}{\makecell[c]{Percentage of\\Zero Counts}} 
  & Person 1 & 2.10\% & 0.14\% & 0.47\% & 0.07\% & 2.95\% \\
  & Person 2 & 9.45\% & 2.16\% & 7.75\% & 0.83\% & 6.90\% \\
\hline
\end{tabular}
\caption{Proportion of zero counts among the abundant genera data in Table~\ref{Data_des} for each individual and body site. }
\label{Data_zero}
\end{table}

\subsection{Do Different Genera Share a Common Mean Reversion Rate?}
In microbiome research, it is often assumed that the mean
reversion rates for different genera are either identical or closely
related~\citep{lloyd2017strains, laitinen2018hierarchical}. This
assumption simplifies modeling and reduces the number of parameters to
estimate. However, the biological plausibility of this assumption
remains a question, especially in complex and heterogeneous microbial
environments such as the human gut.

To evaluate this assumption, we adopted two approaches: profile
likelihood confidence intervals constructed under the OUSSM, and a
likelihood ratio test based on joint modelling of multiple genera.

\subsubsection{Approach 1: Profile Likelihood Confidence Intervals}
\begin{figure}[h]
\centering \includegraphics[scale = 0.25]{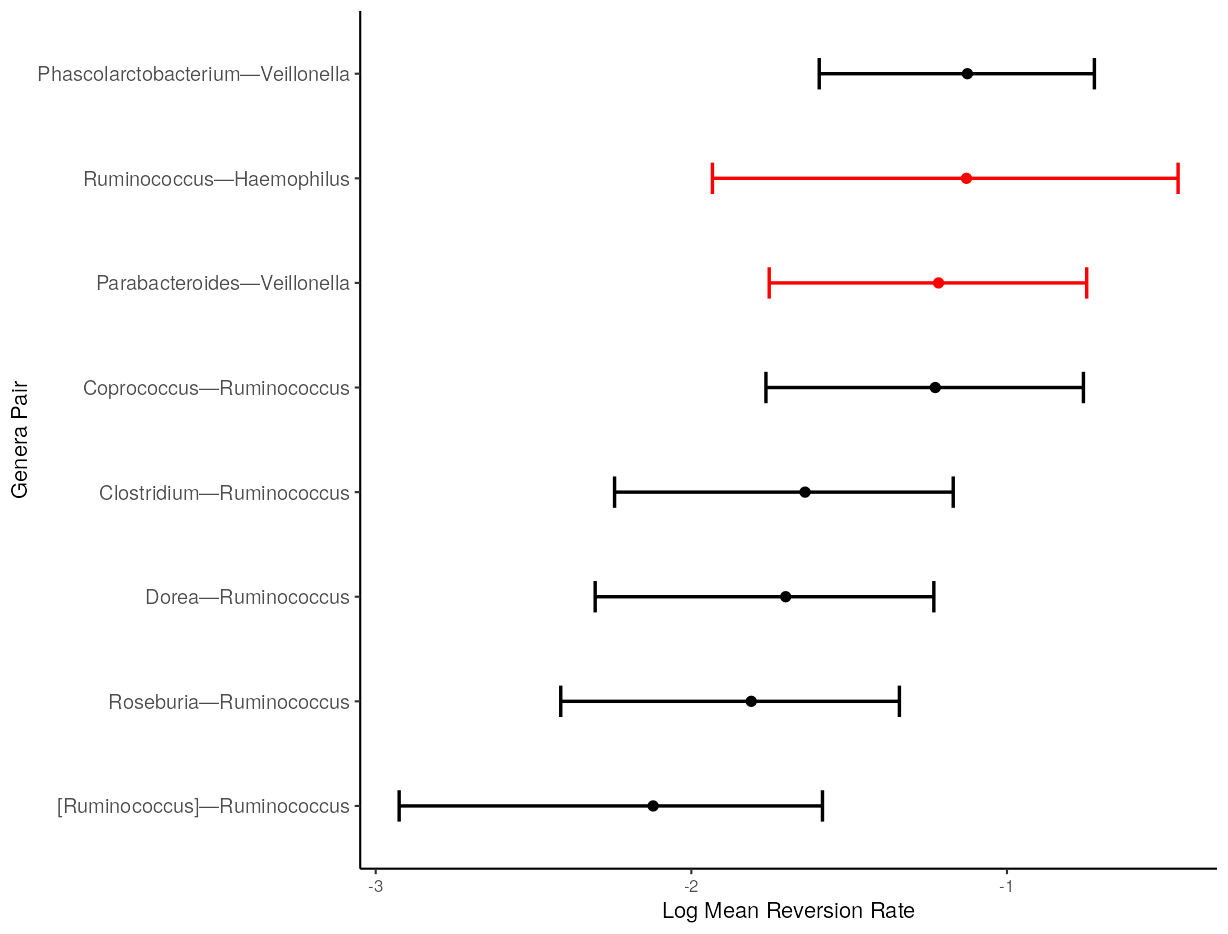}
\caption[95\% Profile Likelihood CIs (David, gut1)]{95\% profile
  likelihood confidence intervals of $\log(\theta)$ for log-ratio data
  from the gut of subject 1 in the dataset from
  \citet{david2014diet}. Black bars represent genus pairs from the
  same phylum; red bars are from different phyla.}
\label{D1CI}
\end{figure}

For each dataset, we first estimated the measurement
error variance matrix $\hat H$ using the method described in
Section~\ref{H_estimate}. We then performed MLE estimation of the
other OUSSM parameters for each log-ratio, treating $\hat H$ as fixed,
and constructed profile likelihood confidence intervals for genus-pair
mean reversion rates. Figure~\ref{D1CI} displays results for Subject 1
from the dataset of \citet{david2014diet}, including only genus pairs
that passed the normality test, the Ljung--Box test for independence,
and the likelihood ratio test of the OUSSM against the local-level
model.

These confidence intervals reveal substantial heterogeneity in
mean-reversion rates across genus pairs. While some estimates indicate
rapid mean reversion, others suggest much slower dynamics. Notably,
genera from the same phylum do not consistently exhibit similar
reversion rates, suggesting that ecological or environmental factors
may play a larger role than taxonomy in determining temporal
stability. These findings are consistent with those of
\citet{kenney2020application}, who also reported considerable
variation in genus-specific mean-reversion rates across body sites.

Similar patterns were observed for the remaining subjects
(Figures~\ref{gut1CI}--\ref{tongue2CI} in Supplementary
Appendix~\ref{AppProfCI}).
Even closely related genera like \textit{Bacteroides} and
\textit{Parabacteroides} may differ substantially in stability. The
right-hand samples of Person~1 in the Moving Picture dataset contain
only eight abundant genera, and only one pair of genera passes the
diagnostic tests, while no genus pair from Subject~2's gut in the
dataset of \citet{david2014diet} passed the diagnostic tests, possibly
because of disruption to the microbiome during the subject’s diarrhea
episode. For these datasets, we provide plots of all CIs, together
with their diagnostic test p-values
(Figures~\ref{rhand1CI}--\ref{D2CI} in Supplementary
Appendix~\ref{AppProfCI}). Because these datasets fail either the
OUSSM vs local-level test or the normality test, the OUSSM may be
misspecified, and the displayed intervals should be interpreted with
caution.

The profile likelihood confidence intervals are presented as
descriptive summaries of uncertainty for individual genus-pair
analyses and are not intended as simultaneous confidence intervals for
all pairs. Consequently, no multiplicity adjustment was
applied. Formal inference regarding heterogeneity in mean-reversion
rates is instead based on the likelihood ratio tests from the joint
modelling framework described below.

\subsubsection{Approach 2: Likelihood Ratio Test based on Joint Modelling of Multiple Genera}

While the pairwise log-ratio analyses provide evidence of
heterogeneity in temporal dynamics, a more direct assessment can be
obtained through a joint model based on the single-genus log-ratio
representation introduced in Section \ref{single}.  Specifically, to
test whether the mean-reversion rates are identical across genera, we
consider the following joint model for the log relative abundances of
the $n_a$ genera:
\begin{eqnarray*}
\log\dfrac{y_{t_n}}{S_{t_n}} &=& x_{t_n} + \mu + \varepsilon_{t_n}, \qquad \varepsilon_{t_n} \sim N(0, \Sigma_\varepsilon), \\
x_{t_{n+1}} &=& e^{-\Theta \Delta t_{n+1}} x_{t_n} + \eta_{t_{n+1}}, \qquad \eta_{t_{n+1}} \sim N(0, Q_{t_{n+1}}),
\end{eqnarray*}
where $y_{t_n}$ denotes the vector of observed OTU counts of $n_a$
genera at time $t_n$, and $x_{t_n}$ is the corresponding vector of
latent log abundances. The matrix $\Sigma_\varepsilon =
\mathrm{diag}(\sigma^2_{\varepsilon,1}, \sigma^2_{\varepsilon,2},
\ldots, \sigma^2_{\varepsilon,n_a})$ represents the taxon-specific
measurement error variances, while $\Theta = \mathrm{diag}(\theta_1,
\theta_2, \ldots, \theta_{n_a})$ contains the mean reversion rates of
the genera. The process noise term $\eta_{t_{n+1}}$ follows a
multivariate normal distribution with covariance $Q_{t_{n+1}} =
\mathrm{diag}\!\left\{\frac{\sigma_1^2}{2\theta_1}\!\left(1-e^{-2\theta_1
  \Delta t_{n+1}}\right), \ldots,
\frac{\sigma_{n_a}^2}{2\theta_{n_a}}\!\left(1-e^{-2\theta_{n_a} \Delta
  t_{n+1}}\right)\right\},$ where $\sigma^2_1, \sigma^2_2, \ldots,
\sigma^2_{n_a}$ are the diffusion variances. For computational
tractability, we assume that the latent dynamics of different genera
evolve independently, so both $\Theta$ and $Q_t$ are
diagonal. Dependencies among genera are captured only indirectly
through the compositional normalization and shared sequencing-depth
effects.

The term $S_{t_n}$ represents the latent sequencing depth, and is
modeled as log-normally distributed with parameters $(\mu_s,
\sigma_s^2)$. Because the mean parameter $\mu_s$ only shifts the
overall scale of $\log\left(\dfrac{y_{t_n}}{S_{t_n}}\right)$ and is
therefore not separately identifiable from the intercept term $\mu$,
we set $\mu_s = 0$ without loss of generality, with $\sigma_s^2$
estimated from the joint modelling.

The likelihood function of this model is evaluated using the Kalman
filter, from which the MLE of all parameters are obtained using the
\texttt{optim} function in \texttt{R} with numerically calculated
derivatives using the \texttt{L-BFGS-B} method.

We compare the null model $H_0: \theta_1 = \cdots = \theta_{n_a}$ with
the alternative $H_A: \text{at least some } \theta_i \text{ differ}$,
using the likelihood ratio test
\[
\Lambda = 2 \big[\ell(\hat{\Theta}_A) - \ell(\hat{\Theta}_0)\big],
\]
where $\ell(\hat{\Theta}_A)$ and $\ell(\hat{\Theta}_0)$ are the
maximized log-likelihoods under the unrestricted and restricted
models, respectively. Since the null model is a special case of the
alternative with one common mean reversion rate, the difference in
dimension is $n_a - 1$.  Because all parameters are estimated in the
interior of the parameter space after log-transformation, standard
likelihood-ratio theory suggests an asymptotic $\chi^2_{n_a - 1}$
distribution.

\begin{table}[ht]
\centering
\renewcommand{\arraystretch}{1.2}
\begin{tabular}{cl|ccccc|c}
\hline
 &  & \multicolumn{5}{c|}{\textbf{Moving picture}} & \textbf{David {\it et al.}} \\
Person &  & Gut & Tongue & R palm & L palm &  &  Gut \\
\hline
\multirow{4}{*}{1} 
  & Test stat & 8.03 & 19.43 & 0.00 & 60.21 & & 3995.92 \\
  & $n_a$ & 12 & 11 & 8 & 31 & & 16 \\
  & $\mathrm{Var}(\mathrm{diag}(\hat{\Theta}_A))$ & 0.2120 & 7.7277 & 2.3777 & 11.0360 & & 0.0194 \\
  & $p$-value & 0.7109 & 0.0351 & 1.0000 & 0.0009 & & 0.0000 \\
\hline
\multirow{4}{*}{2} 
  & Test stat & 178.47 & 153.65 & 208.14 & 91.91 & & 95.56 \\
  & $n_a$ & 32 & 19 & 64 & 30 & & 12 \\
  & $\mathrm{Var}(\mathrm{diag}(\hat{\Theta}_A))$ & 0.0570 & 0.0497 & 0.3989 & 1.1639 & & 0.1813 \\
  & $p$-value & 0.0000 & 0.0000 & 0.0000 & 0.0000 & & 0.0000\\
\hline
\end{tabular}
\caption[LRT for Identical Mean Reversion Rates]{Likelihood ratio test results
  for the null hypothesis of identical mean reversion rates across
  genera in various body sites. Reported are the test statistic,
  number of abundant genera $n_a$, the variance of the estimated mean
  reversion rates $\mathrm{Var}(\mathrm{diag}(\hat{\Theta}_A))$, and
  the corresponding $p$-value.}
\label{tab:LRT_results}
\end{table}

The analyses in Table~\ref{tab:LRT_results} address a global
hypothesis within each body site, namely whether all genera share a
common mean-reversion rate. Because these tests correspond to distinct
biological datasets and are interpreted separately, no multiplicity
correction was applied.  As shown in Table~\ref{tab:LRT_results}, the
null hypothesis is rejected at the 0.05 level in most body sites,
including all samples from the dataset of \citet{david2014diet} and
most regions from the moving picture study. Only the gut and the right
hand of subject~1 exhibit $p$-values above 0.05, suggesting that in
most cases, genera exhibit distinct mean reversion rates. The
estimated variances of the diagonal elements of $\hat{\Theta}_A$
further illustrate this heterogeneity: body sites such as the left
palm and tongue of subject~1 show relatively large variances,
indicating substantial diversity in the strength of mean reversion
among genera, whereas the gut samples from both subjects exhibit
smaller variances, implying more homogeneous temporal dynamics. While
larger $\mathrm{Var}(\mathrm{diag}(\hat{\Theta}))$ values generally
coincide with stronger evidence against $H_0$, the relationship is not
strictly monotonic, as the likelihood ratio test also depends on the
sample size and estimation precision. Together, the variance and the
statistical significance provide strong evidence of substantial
between-genus variation in the rate of mean reversion.

\subsection{Do the Temporal Dynamics Change for the Human Gut after Disruption of the System?}
\begin{figure}[ht!]
\centering
\includegraphics[width=\linewidth]{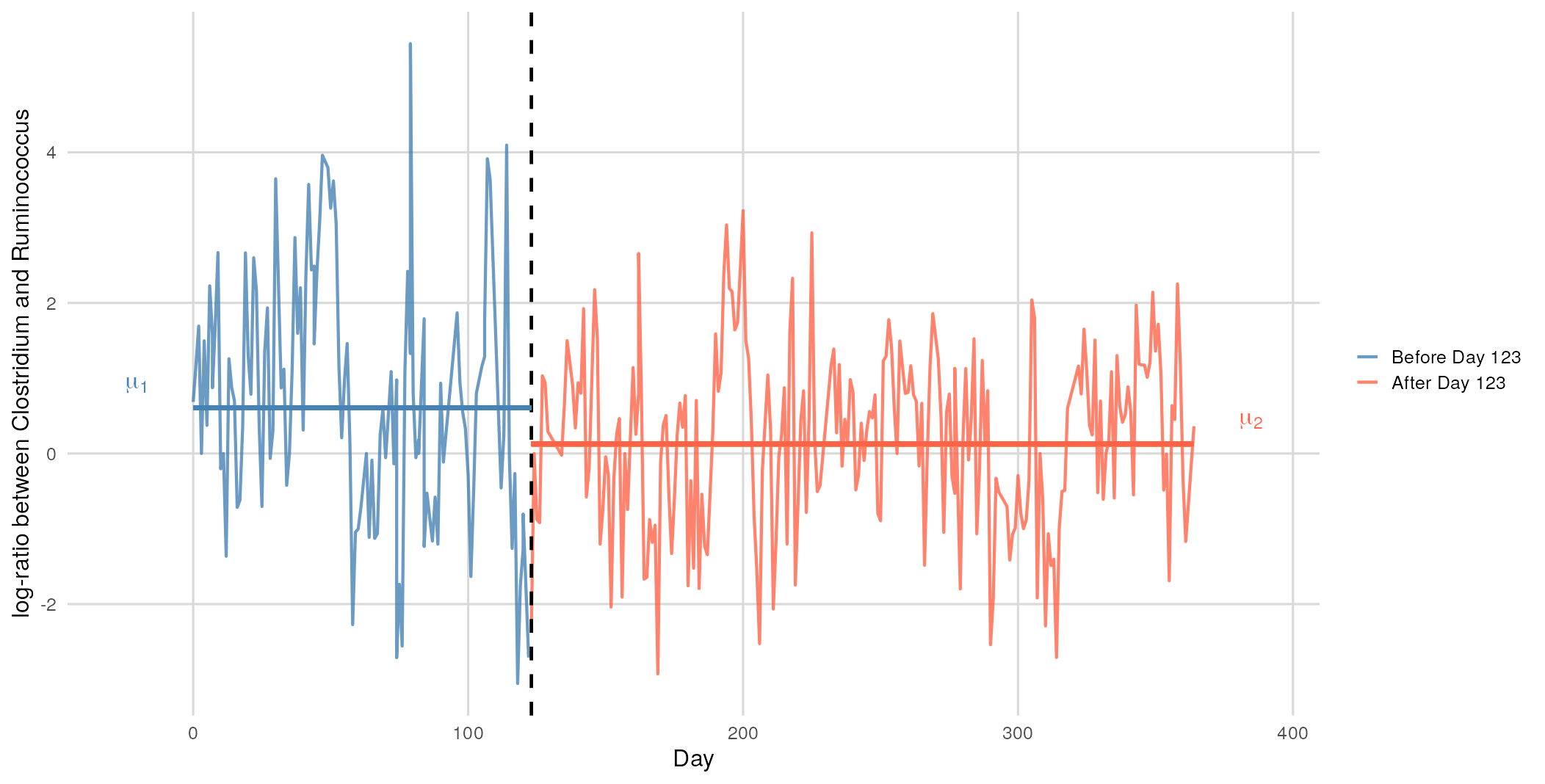}
\caption{Representative log-ratio \textit{Clostridium} and
  \textit{Ruminococcus} for Subject 1 in the dataset of
  \citet{david2014diet}. The dashed vertical line marks Day~123,
  corresponding to the time after a major perturbation event
  experienced by Subject 1.  }
\label{fig:David_shift}
\end{figure}

In the dataset of \citet{david2014diet}, both individuals experienced
notable external disruptions to their microbiomes. Subject A had
several travel events and two episodes of diarrhea, while Subject B
suffered from a period of food poisoning. These events led to
observable shifts in microbiome composition. For example, the
log-ratio of \textit{Clostridium} and \textit{Ruminococcus} for
Subject~1 (Figure~\ref{fig:David_shift}) shows a clear discontinuity
around Day~123.
 
\subsubsection{Statistical framework}\label{StatFramework}
 
Under the OU model, such perturbations are theoretically followed by a
return to equilibrium via the mean reversion process. This suggests
that transient disruptions should have limited long-term influence on
the estimation of temporal dynamics. However, it is clear that in many
cases the mean towards which the process reverts has changed after the
disturbances. It is natural to ask whether the other parameters have
also changed. The time series for Subject~A are split into two
segments: Days~1--123 and Days~124--364. For Subject~B, the data are
divided into Days~1-150 and Days~160-318.
 
The question is whether we should estimate a single set of parameters
$(\hat{\theta}, \hat{\sigma})$ for the full time series while allowing
for different means $(\hat{\mu}_1, \hat{\mu}_2)$ before and after the
disruption, or should we instead estimate separate dynamic parameters
for each segment, i.e., $(\hat{\theta}_1, \hat{\sigma}_1,
\hat{\mu}_1)$ for the pre-disruption period and $(\hat{\theta}_2,
\hat{\sigma}_2, \hat{\mu}_2)$ for the post-disruption period?
 
To address this question, we compare the fit of the combined model
versus the split model using the log-likelihood ratio. We begin by
estimating the measurement error variance matrix $\hat{H}$ as
described in Section~\ref{H_estimate}. Fixing $\hat{H}$, we obtain the
MLE of parameters under each model. The difference in log-likelihoods
is then used as the test statistic for model comparison. Specifically:

\begin{description}

\item[Combined Model (Null Hypothesis $H_0$)]
  Assumes shared dynamics across the entire time series, but allows
  for a different mean before and after the disruption:
    $$\theta, \sigma, \mu_1, \mu_2$$

\item[Split Model (Alternative Hypothesis $H_a$)] Allows
  different dynamic and mean parameters for each segment:
$$
    \theta_1, \sigma^2_1, \mu_1 \quad \text{for } t \leq t_{\text{break}}, \qquad 
    \theta_2, \sigma^2_2, \mu_2 \quad \text{for } t > t_{\text{break}}
$$
\end{description}

To ensure valid inference, we restrict attention to genus pairs that
satisfy the diagnostics introduced in
Section~\ref{diag_test}. Specifically, a pair is included only if it
passes the normality test, the Ljung--Box test for independence, and
the likelihood ratio test comparing the local-level model against the
OUSSM, either for the entire time series or for both pre- and
post-disruption periods.

Through simulation studies in Supplementary Appendix~\ref{sim_split},
we verified that the log-likelihood ratio statistic under the null
hypothesis approximately follows a $\chi^2_2$
distribution. Consequently, p-values for comparing the combined and
split models can be obtained directly from this reference
distribution.

\subsubsection{Likelihood ratio test results}

Table~\ref{tab:LRT_disruption} presents the likelihood ratio test
results comparing the combined and split models of temporal dynamics
before and after documented disruption events. For both individuals in
the dataset of \citet{david2014diet}, most genus pairs favored the
combined model, indicating that the mean-reversion dynamics were
largely preserved across the pre- and post-disruption periods. Only a
small number of genus pairs exhibited significant evidence in favor of
the split model at the 0.05 level.

\begin{table}[ht]
\centering
\renewcommand{\arraystretch}{1.2}
\begin{tabular}{llcccc}
\hline
\textbf{Sub} & \textbf{Genus 1} & \textbf{Genus 2} & \textbf{Test stat} & \textbf{$p$-value} & \textbf{Model} \\
\hline
\multirow{9}{*}{1} 
  & {\it Parabacteroides} & {\it Veillonella  }       & 0.37  & 0.8321 & Combined \\
  & {\it Clostridium}     & {\it Ruminococcus  }      & 4.90  & 0.0862 & Combined \\
  & {\it Coprococcus}     & {\it Ruminococcus }       & 0.15  & 0.9280 & Combined \\
  & {\it Dorea}           & {\it Ruminococcus }       & 2.25  & 0.3242 & Combined \\
  & {\it Roseburia}       & {\it Ruminococcus }       & 5.36  & 0.0687 & Combined \\
  & {\it Ruminococcus}    & {\it Ruminococcus }     & 5.34  & 0.0693 & Combined \\
  & {\it Ruminococcus }   & {\it Haemophilus}         & 63.03 & 0.0000 & Split \\
  & {\it Phascolarctobacterium} & {\it Veillonella}   & 2.29  & 0.3184 & Combined \\
  & {\it Phascolarctobacterium} & {\it Haemophilus}   & 42.85 & 0.0000 & Split \\
\hline
\multirow{1}{*}{2} 
  & {\it Streptococcus}   & {\it Coprococcus}         & 1.84  & 0.3991 & Combined \\
\hline
\end{tabular}
\caption{ Likelihood ratio test results comparing combined versus
  split models of temporal dynamics in the gut microbiome dataset of
  \citet{david2014diet}. Reported are the likelihood ratio test
  statistic, the unadjusted p-value, and the selected model (Combined
  if the split model is not supported; Split otherwise). Note that the
  two \textit{Ruminococcus} entries correspond to distinct lineages
  belonging to the families \textit{Lachnospiraceae} and
  \textit{Ruminococcaceae}, respectively.}
\label{tab:LRT_disruption}
\end{table}

Because these analyses involve multiple genus-pair comparisons within
the same dataset, we additionally applied the Benjamini--Hochberg
procedure to control the false discovery rate at 0.05. The conclusions
were unchanged after adjustment, with the same two genus pairs
involving \textit{Haemophilus} remaining significant. This suggests
that the detected disruption effects are unlikely to be attributable
to multiple-testing artifacts.

Overall, these results indicate that despite a clear change in
equilibrium abundance levels after the documented disruptions, most
mean reversion rates and stochastic variances remained largely
stable. Two genus pairs, both involving genus {\em Hemophilus},
exhibited clear changes in the rate of mean reversion and the
stochastic variance following the disruption, indicating that at
least some aspects of the ecological regulation may have changed after
the disruption.

\subsubsection{Biological interpretation}

\begin{table}[ht]
\centering
\renewcommand{\arraystretch}{1.2}
\begin{tabular}{l|cc}
\hline
  & \multicolumn{2}{c}{\textbf{Genus 1 vs. {\it Haemophilus }}} \\
 Parameter & {\it Ruminococcus } & {\it Phascolarctobacterium} \\
\hline
 {$\theta_1$} (Before day 123) & 0.1139 & 0.7651 \\
 {$\theta_2$} (After day 123)  & 0.0311 & 0.0219 \\
 {$\sigma^2_1$} (Before day 123) & 0.4187 & 2.8736 \\
 {$\sigma^2_2$} (After day 123)  & 0.0579 & 0.0250 \\
 {$\mu_1$} (Before day 123) & 2.3004 & 3.3911 \\
 {$\mu_2$} (After day 123)  & 0.7843 & 1.2428 \\
\hline
\end{tabular}
\caption{Estimated parameters for genus pairs supporting the split
  model in the gut microbiome dataset of \citet{david2014diet}. The
  split model allows distinct mean-reversion rates ($\theta_1$,
  $\theta_2$), diffusion variances ($\sigma^2_1$, $\sigma^2_2$), and
  equilibrium means ($\mu_1$, $\mu_2$) before and after the disruption
  at Day~123.}
\label{tab:disruption_theta}
\end{table}

Table~\ref{tab:disruption_theta} summarizes the estimated
mean-reversion rates ($\theta_1$ and $\theta_2$) for the genus pairs
that supported the split model in Subject~1, who experienced a
travel-associated perturbation in the dataset of
\citet{david2014diet}. The consistent reduction in $\theta$ following
the split indicates a substantial weakening of regulatory strength in
microbial dynamics after the disruption. Biologically, lower
post-disruption mean-reversion rates imply a slower return to
equilibrium following disturbances, consistent with a temporary loss
of homeostatic control within the gut ecosystem. However, we also note
that although the mean-reversion rate drops substantially for both
genera, this is accompanied by an even larger decline in stochastic
variance, meaning that the fitted model has lower marginal variance
for the log-ratio of the two genera in the post-disruption
period. That is, the marginal abundance ratio between the two genera
is more stable after the disruption, but the stabilising forces act on
a slower time-scale. It is plausible that the abundance ratios are
controlled by multiple systems acting on different time-scales, and
that after the disruption, the systems driving the volatile but
quickly stabilising behaviour became substantially less active,
leading the model to fit the remaining low volatility system with
slower mean-reversion. Unfortunately, there is insufficient data to
fit a model incorporating behaviour on multiple time-scales. We cannot
determine whether the system fitted after the disruption represents a
new type of dynamics, or a system that was already present, but masked
by the more volatile system.

For the \textit{Phascolarctobacterium--Haemophilus} pair, the
pronounced decline in the mean-reversion rate could be explained by a
weakening of succinate-mediated regulatory
feedback. \textit{Phascolarctobacterium} is a major succinate consumer
in the human gut and rapidly converts succinate to
propionate~\citep{anthamatten2024stratification}, whereas
\textit{Haemophilus} is not known to efficiently utilize succinate and
has been associated with adverse inflammatory and clinical
outcomes~\citep{zhu2021association}. The observed reduction in mean
reversion therefore may reflect diminished abundance or activity of
\textit{Phascolarctobacterium}, coupled with increased prominence of
\textit{Haemophilus}, leading to less efficient succinate clearance
and weakened metabolic feedback. Such a loss of regulatory control
would slow the restoration of metabolic equilibrium following
perturbations, consistent with the substantially lower post-disruption
mean-reversion rate estimated by the OUSSM.

A comparable pattern is evident for the
\textit{Ruminococcus--Haemophilus} pair, although the relevant
metabolic processes differ. \textit{Ruminococcus} species are
established commensals that degrade complex carbohydrates and produce
short-chain fatty acids that support epithelial health and microbial
stability~\citep{la2018ruminococci,
  crost2023ruminococcus}. \textit{Haemophilus}, by contrast, lacks a
prominent role in carbohydrate fermentation or short-chain fatty acid
production and is more commonly associated with dysregulated or
opportunistic microbial states, including associations with
neuropsychiatric and respiratory
conditions~\citep{zhu2021association}. The reduction in mean-reversion
rate for this pair therefore plausibly reflects diminished
fermentative capacity and weakened feedback control in the
post-disruption period.

Taken together, these findings suggest that the stabilizing feedback
mechanisms of the gut microbiome may be subject to long-term effects
from short-term perturbations, resulting in new temporal dynamics for
metabolically important genera such as \textit{Phascolarctobacterium}
and \textit{Ruminococcus}. The lower post-disruption mean-reversion
rates, paired with reduced short-term volatility suggest that processes
that previously drove short-term fluctuation in microbial composition
were substantially reduced after the perturbation. 

\section{Conclusion and Future Work}
\label{conclusion}

This study presents a rigorous likelihood-based framework for modeling
and analyzing longitudinal microbiome data using Ornstein--Uhlenbeck
(OU) state-space models. By focusing on log-ratio transformed
microbial abundance data, we address the compositional nature of
sequencing outputs and provide interpretable summaries of microbial
temporal behavior.

We demonstrate the utility of the OU-based model in capturing
mean-reverting dynamics and introduce profile likelihood methods to
construct confidence intervals for the mean reversion rate. This
inferential approach quantifies parameter uncertainty and provides
insights into the temporal stability of microbial communities. Using a
likelihood ratio test, we showed that different genera within a
community can exhibit distinct temporal dynamics and, in particular,
different mean-reversion rates. This possibility was previously
suggested by \citet{kenney2020application}, but their analysis did not
account for measurement error.

We further developed a likelihood ratio test for assessing
disruption-induced changes in temporal dynamics. For the longitudinal
dataset of \citet{david2014diet}, there was little evidence of
disruption-induced changes for most genus pairs: although equilibrium
abundance levels often shifted, the estimated mean-reversion rates and
diffusion variances generally remained stable. However, a small number
of genus pairs, particularly those involving \textit{Haemophilus},
exhibited substantial reductions in mean-reversion rates and diffusion
variances following disruption. These findings could represent a
weakening of stabilizing feedback mechanisms within the microbial
community, consistent with dysbiosis, or they could represent a
reduction in processes driving short-term fluctuation.

Several directions for future research remain. The current framework
assumes Gaussian noise and linear mean-reverting dynamics and does not
explicitly model ecological interactions among taxa. Extensions to
multivariate state-space models with interaction structures may
provide a richer description of community dynamics, although such
models present substantial computational and inferential
challenges. The current framework treats estimated measurement-error
variances as fixed in the second stage. Incorporating first-stage
uncertainty into downstream inference represents an important
direction for future work. It would also be valuable to develop
methods for detecting disruption times when changepoints are unknown,
to improve inference for sparse compositional data with many zero
counts, and to incorporate covariates, hierarchical structures across
individuals, or more flexible temporal processes beyond the OU
model. Such extensions may further enhance the applicability of
state-space models for understanding microbial community dynamics.

\section*{Supplementary information}

Supplemenary Appendix~\ref{sim_split} performs a simulation study to
confirm that the finite-sample log-likelihood ratio statistic from
Section~\ref{StatFramework} is well approximated by a chi-squared
distribution. Supplemenary Appendix~\ref{AppProfCI} shows confidence
intervals for $\theta$ for log-ratio data for the datasets not shown
in Figure~\ref{D1CI}.

\section*{Declarations}

\subsection*{ Funding}

The authors gratefully acknowledge support from NSERC Discovery Grants
RGPIN-2023-03238 and RGPIN-2023-03332.

\subsection*{ Competing interests}

The authors declare that they have no known competing financial
interests or personal relationships that could have appeared to
influence the work reported in this paper.

\subsection*{ Ethics approval and consent to participate}

Not applicable

\subsection*{ Consent for publication}

Not applicable

\subsection*{ Data availability }

The data are publicly available from the papers~\citep{david2014diet}
and \citep{caporaso2011moving}. 

\subsection*{ Materials availability}

Not applicable

\subsection*{ Code availability }

The code is available as an R package at
\url{https://github.com/ShanglunLi/OUSSM1d}.

\subsection*{ Author contributions}

All authors contributed to the study design, methodology development,
interpretation of results, and revision of the manuscript. SL
implemented the analysis methods and compiled the results. All authors
read and approved the final version of the paper.

\bibliographystyle{plainnat}
\bibliography{reference}

\end{document}


\title{Supplementary Appendices to ``State Space Modeling of the Ornstein--Uhlenbeck Process with Measurement Error: An Application to Microbiome Data''}
\maketitle

\appendix

\section{Simulation Study of Combined and Split Model}
\label{sim_split}
\begin{figure}[h]
\centering
\includegraphics[scale = 0.7]{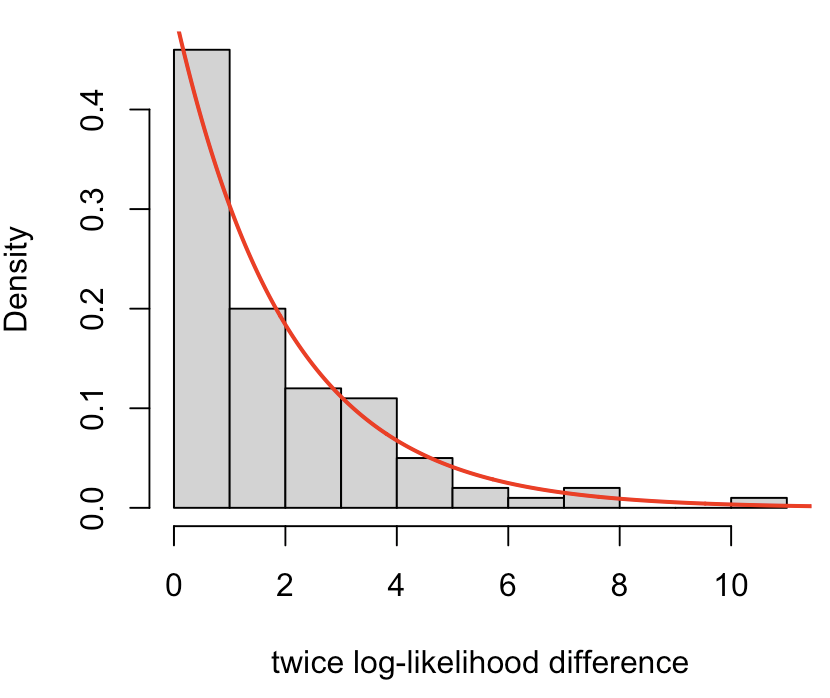}
\caption{Histogram of twice log-likelihood differences between the
  segmented and combined models based on 100 simulations under the
  null hypothesis. The red curve shows the $\chi^2$ distribution with
  2 degrees of freedom. The alignment suggests that the likelihood
  ratio statistic under the null approximately follows a $\chi^2_2$
  distribution as suggested by statistical theory.}
\label{hist_L_diff}
\end{figure}

To validate the null distribution of the likelihood ratio statistic,
we simulated 100 datasets under the combined model, where a single set
of dynamic parameters $(\hat{\theta}, \hat{\sigma})$ governs the
entire time series, but the mean levels $(\hat{\mu}_1, \hat{\mu}_2)$
are allowed to differ before and after the disruption. This setup
reflects the null hypothesis in which the dynamics remain unchanged
across the disruption. Each simulated dataset was then fit with both
the combined and split models, and the resulting log-likelihood
differences were computed. As shown in Figure~\ref{hist_L_diff}, the
empirical distribution of these differences closely aligns with a
chi-squared distribution with 2 degrees of freedom, validating the use
of this distribution for hypothesis testing.

\newpage
\section{Profile Confidence Interval}
\label{AppProfCI}

\begin{figure}[h]
\centering
\includegraphics[scale = 0.3]{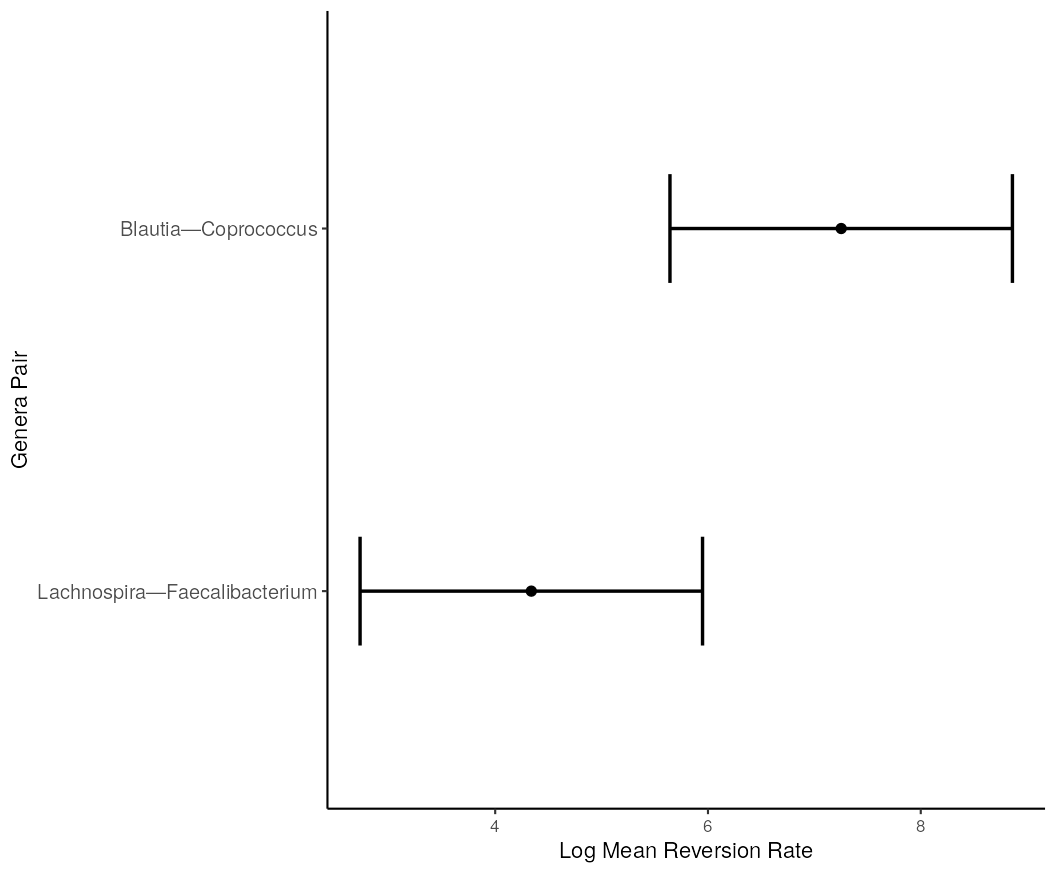}
\caption{95\% profile likelihood confidence intervals for the
  log-ratio data from the gut of subject 1 in the moving picture
  dataset.  }
\label{gut1CI}
\end{figure}

\begin{figure}[h]
\centering
\includegraphics[scale = 0.3]{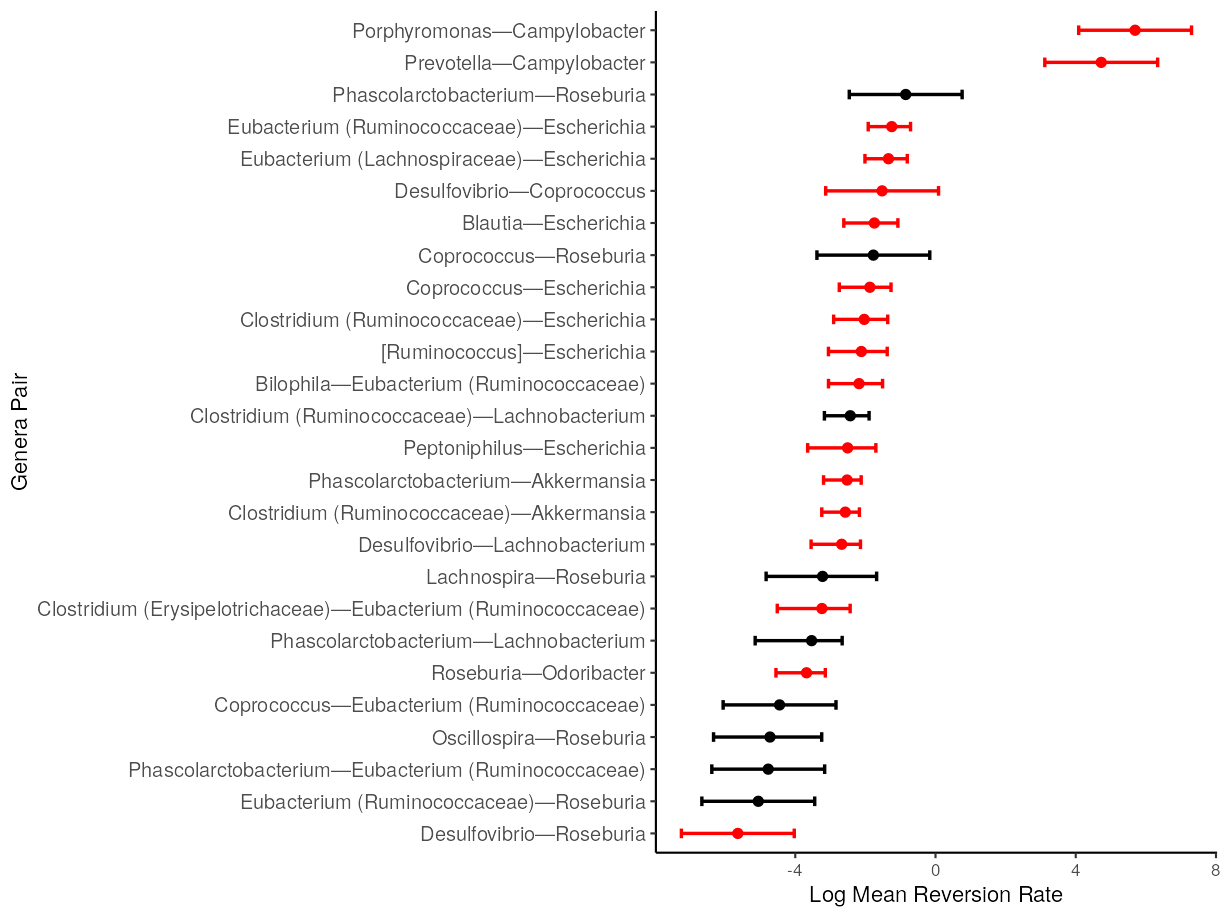}
\caption{95\% profile likelihood confidence intervals for the
  log-ratio data from the gut of subject 2 in the moving picture
  dataset.  }
\label{gut2CI}
\end{figure}

\begin{figure}[h]
\centering
\includegraphics[scale = 0.3]{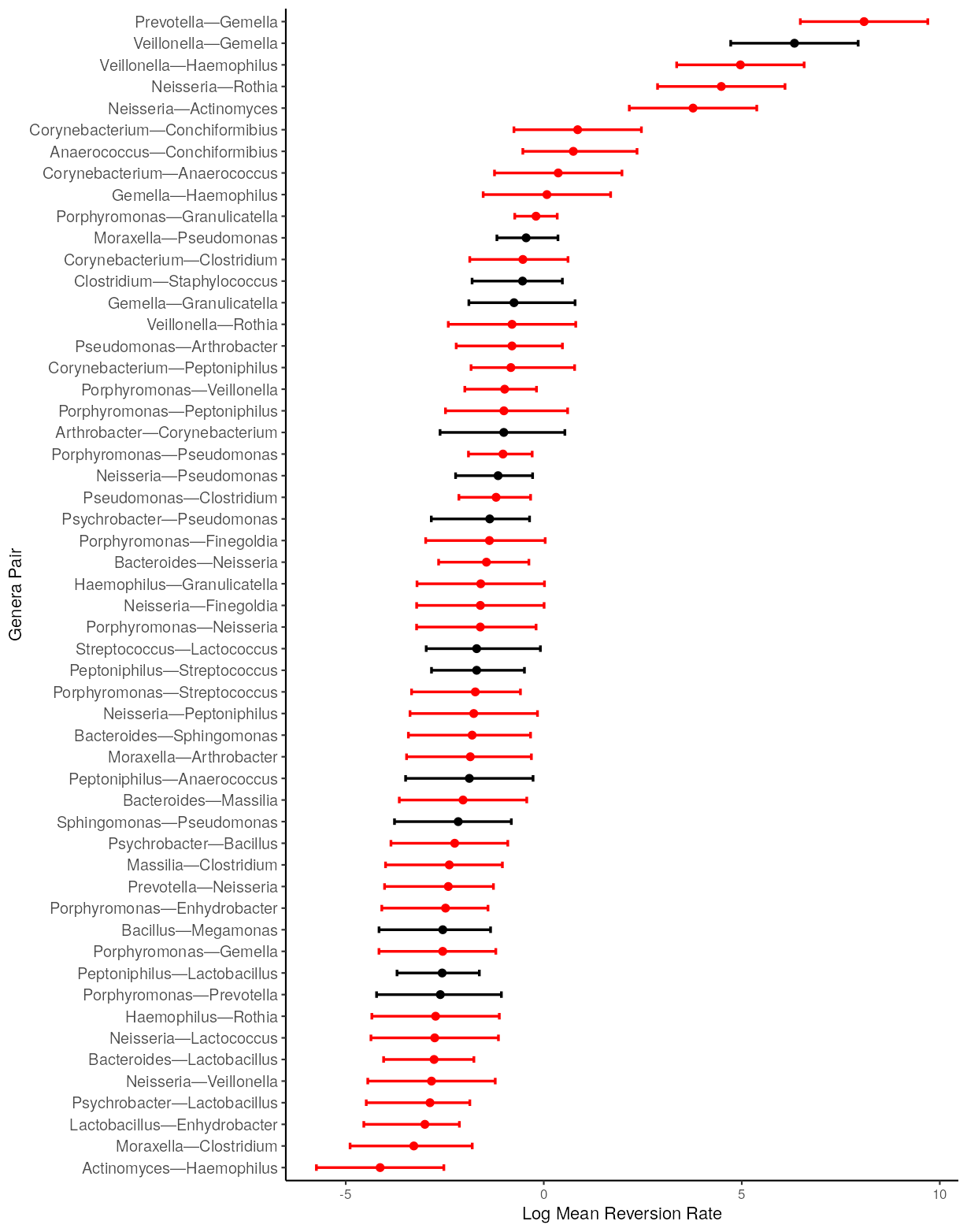}
\caption{95\% profile likelihood confidence intervals for the
  log-ratio data from the left hand of subject 1 in the moving picture
  dataset.  }
\label{lhand1CI}
\end{figure}

\begin{figure}[h]
\centering
\includegraphics[scale = 0.3]{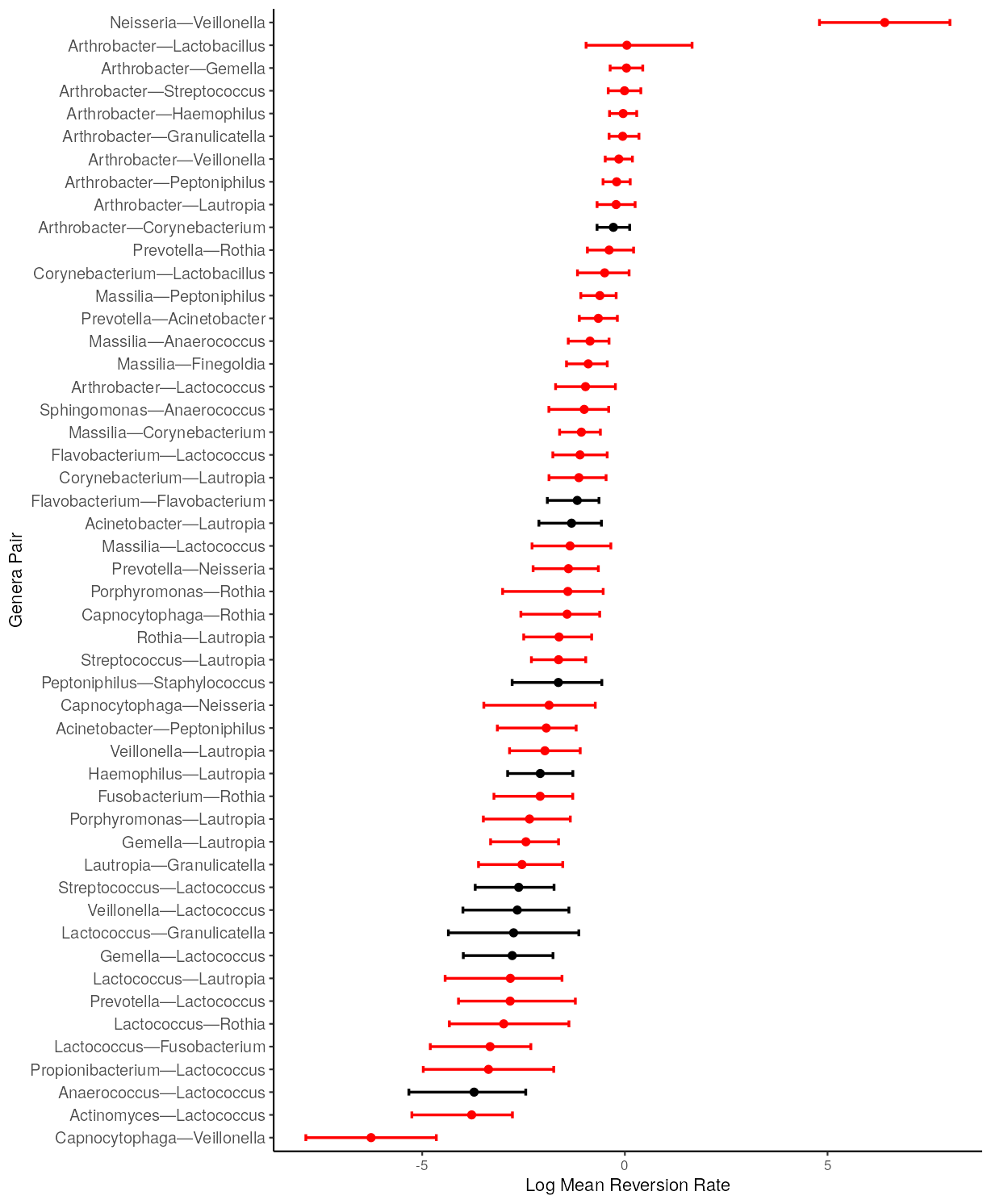}
\caption{95\% profile likelihood confidence intervals for the
  log-ratio data from the left hand of subject 2 in the moving picture
  dataset.  }
\label{lhand2CI}
\end{figure}

\begin{figure}[h]
\centering
\includegraphics[scale = 0.3]{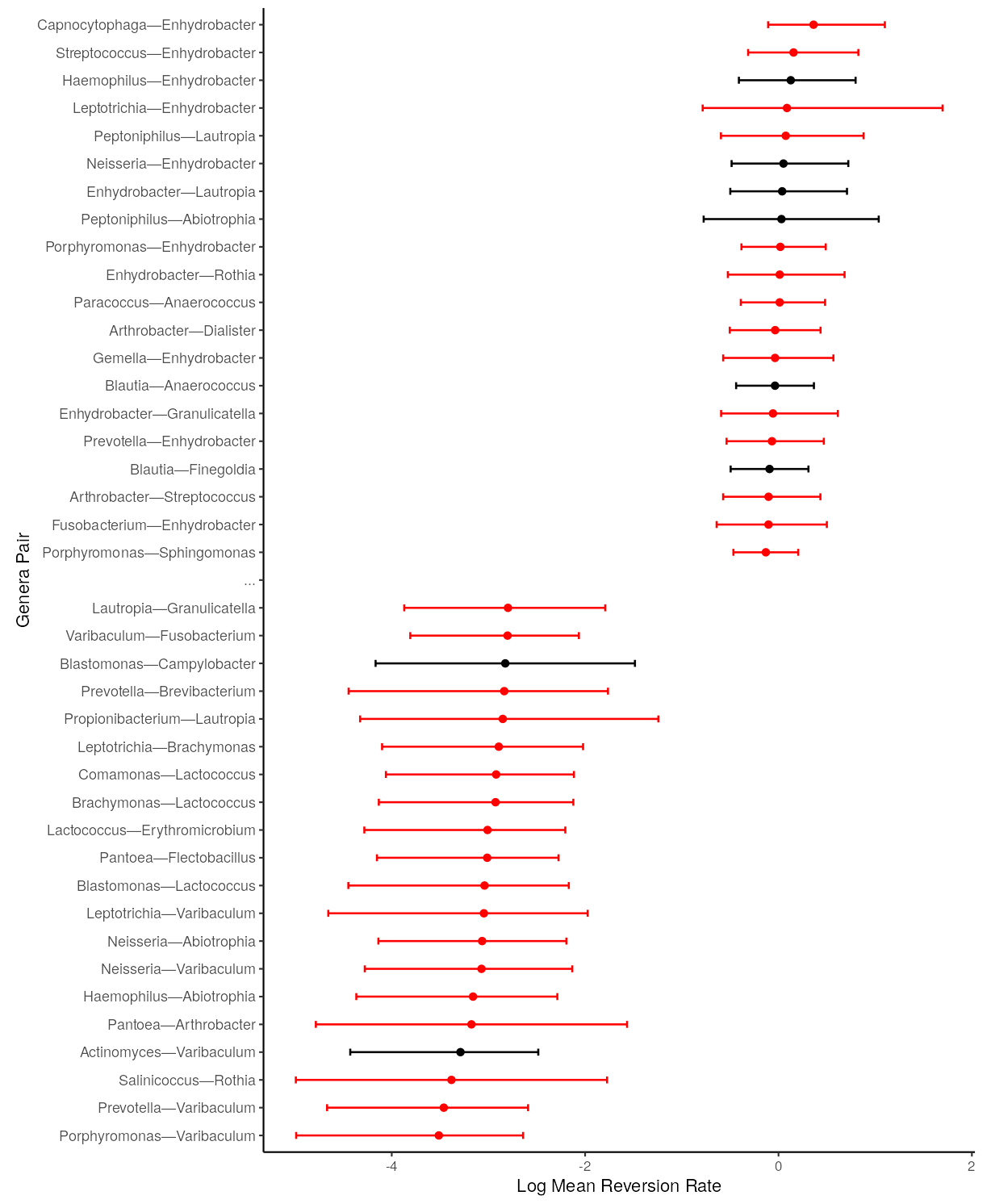}
\caption{95\% profile likelihood confidence intervals for the
  log-ratio data from the right hand of subject 2 in the moving
  picture dataset. Only the top 20 CIs and bottom 20 CIs are shown.  }
\label{rhand2CI}
\end{figure}

\begin{figure}[h]
\centering
\includegraphics[scale = 0.3]{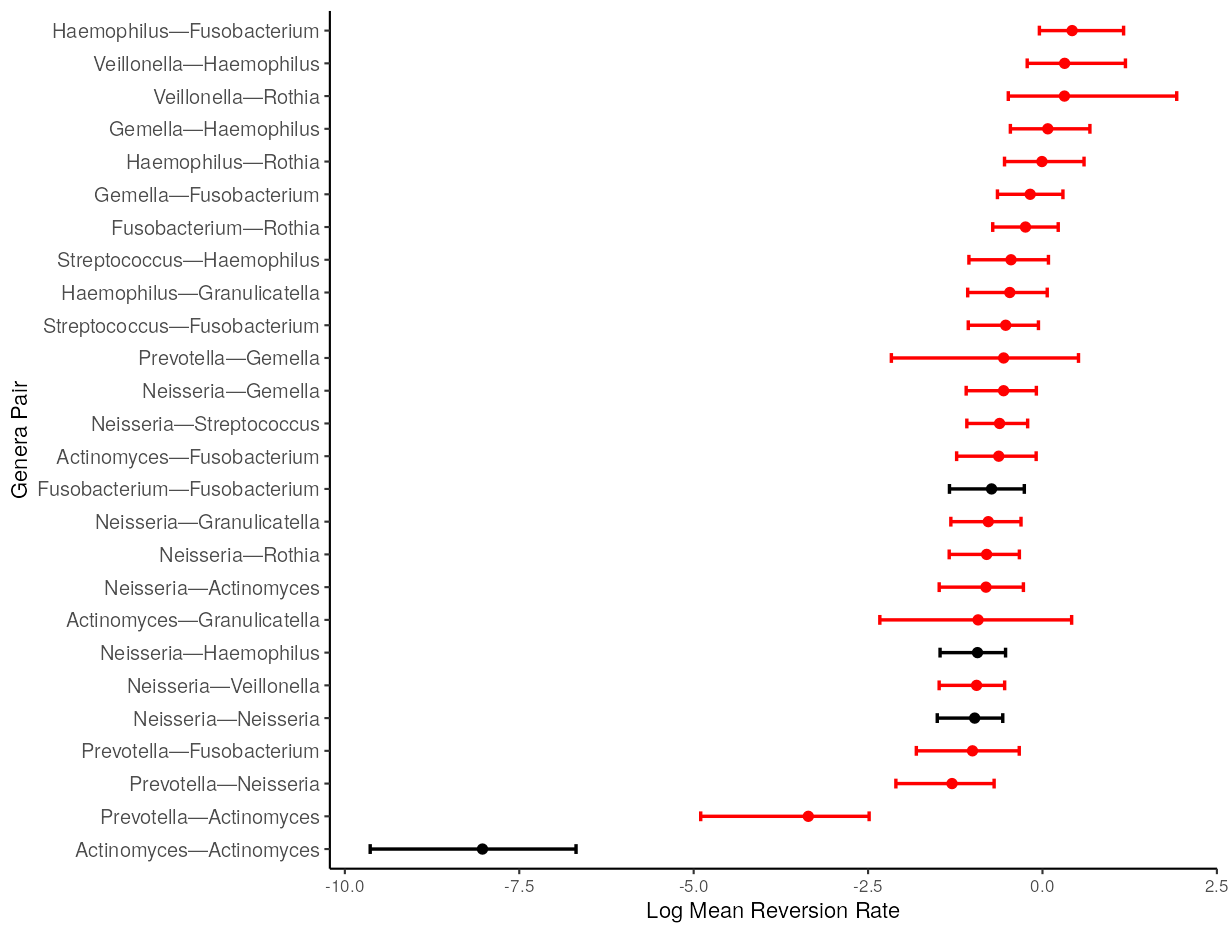}
\caption{95\% profile likelihood confidence intervals for the
  log-ratio data from the tongue of subject 1 in the moving picture
  dataset.  }
\label{tongue1CI}
\end{figure}

\begin{figure}[h]
\centering
\includegraphics[scale = 0.3]{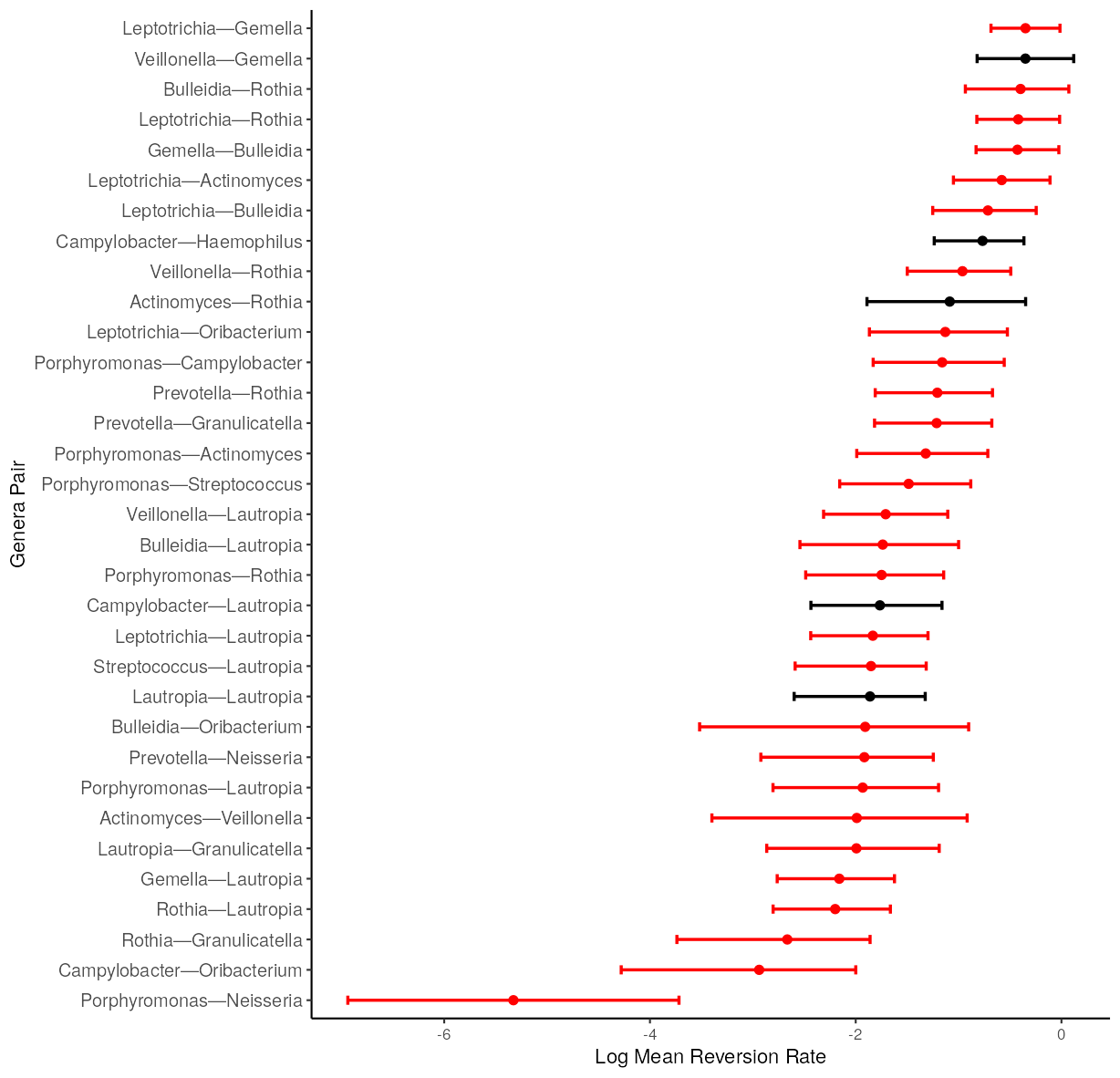}
\caption{95\% profile likelihood confidence intervals for the
  log-ratio data from the tongue of subject 2 in the moving picture
  dataset.  }
\label{tongue2CI}
\end{figure}

\begin{figure}[h]
\centering
\begin{adjustwidth}{-1.5cm}{-1.5cm}
\includegraphics[scale = 0.26]{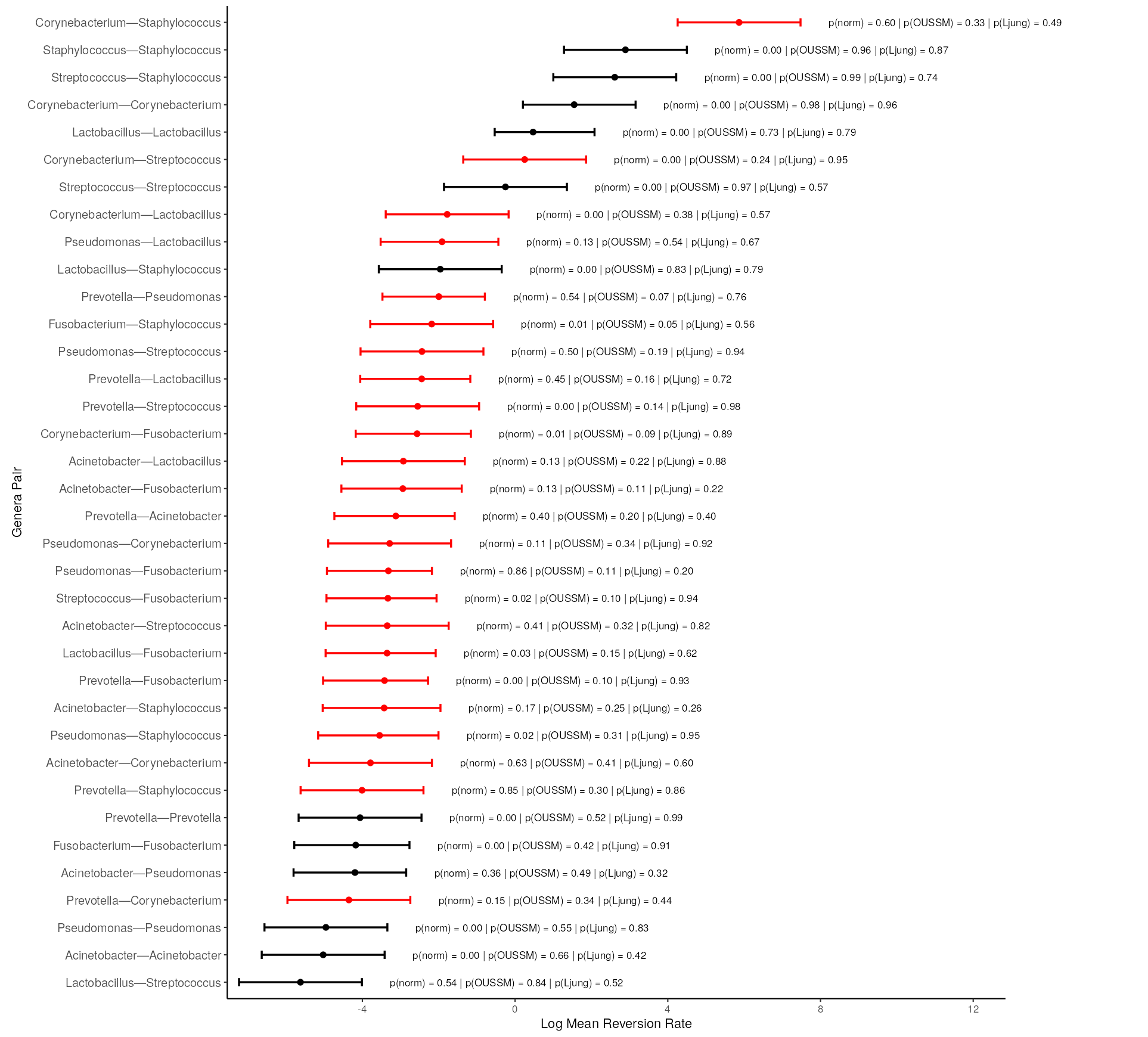}
\end{adjustwidth}
\caption{95\% profile likelihood confidence intervals for the
  log-ratio data from the right hand of subject 1 in the moving
  picture dataset.  }
\label{rhand1CI}
Black bars represent genus pairs from the same phylum; red bars are
from different phyla. Each bar is annotated with the $p$-values of
goodness-of-fit tests for the OUSSM: \texttt{p(norm)} is the $p$-value
of a normality test for the standardised residuals, so a rejection
indicates that the OUSSM is not an adequate model. \texttt{p(OUSSM)}
is the $p$-value of a likelihood-ratio test between a local-level
model and an OUSSM, so a rejection indicates that the OU process is a
better fit for the dynamics of the latent process than Brownian
motion. \texttt{p(Ljung)} is the $p$-value of a Ljung-Box test for
serial correlation in the residuals, so a rejection indicates that
there is unmodelled autocorrelation between time-points.
\end{figure}

\begin{figure}[h]
\centering
\begin{adjustwidth}{-1.5cm}{-1.5cm}
\includegraphics[scale = 0.24]{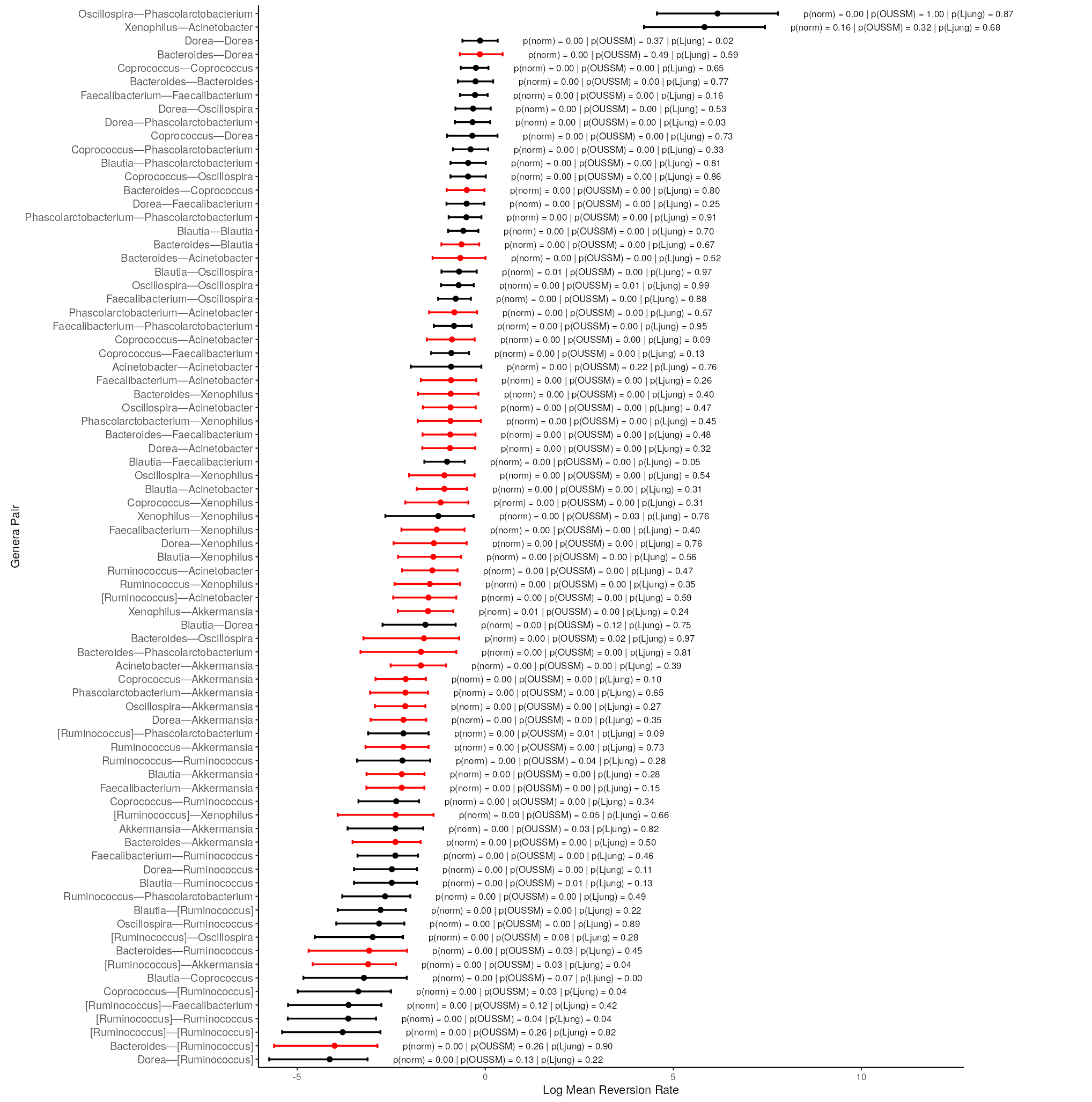}
\end{adjustwidth}
\caption{95\% profile likelihood confidence intervals for log-ratio
  data from the gut of subject 1 in the David {\it et
    al}\ dataset~\cite{david2014diet}.}  Black bars represent genus
pairs from the same phylum; red bars are from different phyla. Each
bar is annotated with the $p$-values of goodness-of-fit tests for the
OUSSM: \texttt{p(norm)} is the $p$-value of a normality test for the
standardised residuals, so a rejection indicates that the OUSSM is not
an adequate model. \texttt{p(OUSSM)} is the $p$-value of a
likelihood-ratio test between a local-level model and an OUSSM, so a
rejection indicates that the OU process is a better fit for the
dynamics of the latent process than Brownian motion. \texttt{p(Ljung)}
is the $p$-value of a Ljung-Box test for serial correlation in the
residuals, so a rejection indicates that there is unmodelled
autocorrelation between time-points.
\label{D2CI}
\end{figure}

\bibliographystyle{plainnat}
\bibliography{reference}